%% file: harnessllm.tex
\documentclass[sigconf]{acmart}

\input{premable}

\begin{document}

\title{\codename: Rust Verification Harness Generation with Large Language Models}


\author{Minghua Wang}
\authornote{Corresponding author.}
\orcid{0000-0002-2270-2076}
\affiliation{%
  \institution{Ant Group}
  \city{Beijing}
  \country{China}}
\email{minghua.wmh@antgroup.com}

\author{Yuwei Liu}
\orcid{0000-0001-5170-3388}
\affiliation{%
  \institution{Ant Group}
  \city{Hangzhou}
  \state{Zhejiang}
  \country{China}}
\email{lyw458372@antgroup.com}

\author{Lin Huang}
\orcid{0009-0002-5659-1471}
\affiliation{%
  \institution{Ant Group}
  \city{Beijing}
  \country{China}}
\email{linyu.hl@antgroup.com}


\input{abs}

% \input{ccs-concepts-keywords}

\maketitle

\input{intro}
\input{bkgd}
\input{overview}

\input{impl}
\input{eva}

\input{discussion}
\input{related}
\input{conclusion}

\bibliographystyle{ACM-Reference-Format}
\bibliography{sample-base}

\end{document}

%% file: premable.tex
\usepackage{graphicx}
\usepackage{xspace} 
\usepackage[ruled,vlined,linesnumbered]{algorithm2e}
\usepackage{float}
\usepackage{enumitem}
\usepackage{listings}
\usepackage{xcolor}
\usepackage{caption}
\usepackage{subcaption}
\usepackage{multirow}
\usepackage{colortbl}
\usepackage{fancyvrb}

\newcommand{\codename}{{\emph{HarnessLLM}}\xspace}
\newcommand{\candidate}{scenario-separated test function}
\newcommand{\candidates}{scenario-separated test functions}
\newcommand{\sketch}{scenario function}
\newcommand{\sketchs}{scenario functions}

\newcommand{\testslineofcode}{6651}
\newcommand{\teststotalnumber}{494} % number of all test cases
\newcommand{\numofsketchs}{294} % sum of #Call Scen. in Table 2.
\newcommand{\numofharns}{294} % same as \numofsketchs
\newcommand{\avgtimegen}{2.42 minutes}
\newcommand{\avgtimegensec}{145 seconds}
\newcommand{\scenpresvrate}{94.66\%}
\newcommand{\numofcrates}{9}
\newcommand{\autoharnsuprate}{41\%}

\newcommand{\textttsmall}[1]{\texttt{\small #1}}
\newcommand{\code}[1]{{\textttsmall{#1}}}

\definecolor{blizzardblue}{rgb}{0.67, 0.9, 0.93}
\definecolor{lightred}{rgb}{1,0.9,0.9}  % 浅红色背景

\newcommand{\verysmall}{\fontsize{6}{6}\selectfont} % font for fig/*.tex
\DeclareCaptionFont{caplabelsmall}{\scriptsize}

\definecolor{lightgray}{gray}{0.95} 
\newsavebox{\zeropaddingboxbox}
\newenvironment{zeropaddingbox}
  {\vspace{0pt}\noindent%
   \begin{lrbox}{\zeropaddingboxbox}%
   \begin{minipage}[t]{\linewidth}\vspace{-0.1pt}%
  }
  {\vspace{-0.1pt}\end{minipage}\end{lrbox}%
   \colorbox{lightgray}{\usebox{\zeropaddingboxbox}}}

\makeatletter
\renewcommand*\sectionautorefname{\S\@gobble}
\def\sectionautorefname{\S\@gobble}
\def\subsectionautorefname{\S\@gobble}
\def\subsubsectionautorefname{\S\@gobble}
\def\chapterautorefname{\S\@gobble}
\makeatother

\input{pygtex/default.pygstyle}

\newenvironment{ScriptSizePygex}[1][\scriptsize]
{
  #1
  \ttfamily
  \begin{minipage}{\linewidth}
}
{
  \end{minipage}
}

%% file: abs.tex
\begin{abstract}
Rust’s ownership and type system offer strong memory safety guarantees, but unsafe code and runtime panics still present significant risks. Formal verification is essential to ensure memory safety, but developing verification harnesses remains a challenging and manual task. Although large language models (LLMs) have shown strong performance in various code analysis tasks, directly applying them to harness generation often results in inaccurate API invocations, inefficient nondeterministic data generation, and fabricated fixes.

In this paper, we present \codename, an automated workflow that leverages LLMs to generate verification harnesses for Rust code directly from existing test suites. \codename automatically extracts calling scenarios from test cases, generates nondeterministic arguments based on dependency analysis, and incrementally synthesizes harnesses. It then iteratively refines the harnesses, preserving critical code regions and reporting fabricated types or functions to LLMs for correction.
In our evaluation on 9 real-world Rust codebases, \codename extracted \numofsketchs{} calling scenarios from \teststotalnumber{} test cases with \scenpresvrate{} precision and generated harnesses for all scenarios in an average of \avgtimegensec{} each. It outperformed the existing approach, Autoharness, which succeeded on only \autoharnsuprate{} of those scenarios. Finally, 6 real-world memory safety bugs were detected using the generated harnesses, demonstrating the practical utility of our approach in verification. To our knowledge, this is the first work to use LLMs for generating harnesses aimed at memory safety verification in real-world Rust projects.
\end{abstract}

%% file: intro.tex
\section{Introduction}
Rust’s ownership and type system provide strong guarantees against memory safety issues. However, unsafe code can bypass these guarantees and introduce vulnerabilities~\cite{mem-safety-challenge, mem-thread-safety, unsafe-interview}. Even safe Rust inserts runtime assertions to prevent overflows and out-of-bounds accesses, which may trigger panics unacceptable in security-critical contexts. While program analysis techniques~\cite{mir-checker, safedrop, yuga, rudra} detect many bugs, they lack sound guarantees. Formal methods like theorem proving~\cite{rustbelt, aeneas, refined-rust, hax, electrolysis} and deductive verification~\cite{verus, prusti, creusot} offer stronger assurances but demand substantial manual effort. Bounded model checking (BMC)~\cite{kani, smack}, as one type of automated approach, has been widely adopted in memory safety verification. It encodes execution traces as SAT/SMT problems and produces counterexamples for violations and bounded proofs.

Verification harnesses are essential for applying BMC but are tedious to craft manually, requiring developer effort to identify meaningful API scenarios. Existing tools like RULF~\cite{rulf} and RPG~\cite{rpg} can synthesize Rust API invocation sequences, but they often lack sufficient semantic insight to prevent API misuse. PropProof~\cite{propproof} converts proptest cases~\cite{proptest} into Kani harnesses but is limited by test availability. Kani’s Autoharness~\cite{kani, autoharness} automates harness generation for functions whose parameter types implement \code{kani::Arbitrary} but fails to produce nondet\footnote{Throughout this paper, the terms nondet (abbreviated) and nondeterministic (full form) are employed synonymously.} user-defined types that do not implement this trait. Traditional program analysis approaches face challenges with Rust’s advanced features (e.g., generics, closures, higher-order functions) and suffer from LLVM and Rustc version incompatibilities.
Recent LLMs have demonstrated superior capabilities at code summarization~\cite{llm-code-sum}, vulnerability detection~\cite{promptfuzz, llm-api-misuse}, software engineering~\cite{llm-apr, llm-se-android}, and program verification~\cite{autoverus, safe}, yet have not been applied to harness generation. This paper introduces an LLM-assisted approach for automatically generating Rust verification harnesses, enabling BMC to check for memory safety violations and runtime panics.

However, directly using LLMs to generate verification harnesses for Rust code presents several challenges. 

\noindent\textbf{C1}: LLMs struggle to capture complete calling scenarios. Formal verification requires validating diverse API usage scenarios to ensure soundness. For small programs, providing the complete code to an LLM may suffice, but in real-world codebases the context window limitations prevent the inclusion of all relevant code.

\noindent\textbf{C2}: LLMs have difficulty generating nondet complex arguments. Synthesizing harnesses requires creating nondet arguments. Although LLMs can produce nondet Rust primitive types (e.g., integers, boolean), constructing nondet complex, interdependent types remains challenging. Without precise and clear instructions, LLMs struggle to resolve dependencies to construct required arguments. 

\noindent\textbf{C3}: LLMs may fabricate types and functions. To produce syntactically correct harnesses, Rust compilers are used to compile generated code and provide error feedback. However, due to hallucinations, LLMs may fabricate type definitions or function implementations to bypass compiler errors after multiple attempts. Moreover, error corrections may include arbitrary modifications or even alter previously correct code, inadvertently introducing new errors. This reduces repair efficiency and makes generating a correct harness within a limited number of iterations more difficult.

We observe that, unlike C/C++ projects, Rust codebases often include high-quality test suites covering diverse API usage. Inspired by this, we propose \codename, a method that leverages LLMs to automatically generate verification harnesses from existing tests.
\codename consists of four phases: code analysis, calling scenario extraction, harness synthesis, and harness compilation. 
In the code analysis phase, we extract types, traits as well as function definitions, and identify test cases related to the target API.
During the calling scenario extraction phase, we isolate individual calling scenarios from the tests (addressing C1) and encapsulate each as an independent ``scenario function.” 
In the harness synthesis phase, we first generate nondet parameters and then invoke scenario functions to produce a complete harness. To generate nondet parameters, we construct a type dependency graph and, based on this, produce Chain-of-Thought (CoT) instructions that guide the LLM to incrementally construct the required data types with their public constructors (addressing C2). 
Finally, in the harness compilation phase, we compile the generated harness with Kani and feed any errors back to the LLM. We specify code regions that must remain unchanged and, at each iteration, check and report any fabricated types or functions to the LLM (addressing C3).

We evaluated \codename{} on \numofcrates{} real-world Rust libraries from crates.io~\cite{crates-io}. From \teststotalnumber{} test cases, \codename{} extracted \numofsketchs{} invocation scenarios with \scenpresvrate{} precision, generated \numofharns{} syntactically correct verification harnesses with a 100\% success rate and an average generation time of \avgtimegensec{} seconds per harness. On the same dataset, Kani’s Autoharness achieved only \autoharnsuprate{} coverage. Ablation studies confirmed the effectiveness of \codename{}’s components, and applying the generated harnesses uncovered 6 real-world memory safety bugs, demonstrating its practical utility. 

We summarize the contribution as follows:
\begin{itemize}[leftmargin=1em]
    \item We introduce \codename{}, the first automated workflow that uses LLMs to generate Rust verification harnesses from existing test suites, enabling memory safety verification of real-world Rust codebases.
    
    \item We decompose harness generation into phases of code analysis, call scenario extraction, harness synthesis, and harness compilation, and design prompting strategies to guide LLMs in producing reliable, high-quality outputs at each stage.

    \item In evaluations across \numofcrates{} real‐world Rust libraries, \codename{} achieved \scenpresvrate{} precision in scenario extraction, a 100\% success rate in harness generation, significantly outperforming Kani’s Autoharness which only generated harnesses for only \autoharnsuprate{} of scenarios, with an average generation time of \avgtimegensec{} per harness.

    \item Applying the generated harnesses uncovered 6 real‐world memory safety bugs (5 fixed), demonstrating \codename{}’s practical utility for Rust memory safety verification.
\end{itemize}

%% file: bkgd.tex
\section{Background}

\noindent\textbf{Safe Rust and Unsafe Rust.}
Rust's popularity stems from its memory safety features and minimal runtime overhead. However, the unsafe keyword allows operations that bypass safety guarantees, introducing memory safety vulnerabilities. Furthermore, even in safe Rust, runtime errors, including panics, can still occur. The compiler inserts assertions in critical statements (e.g., arithmetic operations and unwrapping), which, if violated, cause program aborts. In this paper, we generate harnesses to verify the absence of memory safety issues and runtime panics.

\noindent\textbf{Bounded Model Checking.}
Bounded Model Checking (BMC) is a widely used technique for verifying memory safety in unsafe Rust. It encodes program traces as symbolic SAT/SMT problems and employs solvers to provide bounded proofs. However, BMC requires setting fixed bounds on loop iterations and recursion depths. Small bounds risk incomplete unwinding and may miss genuine bugs, while large bounds can cause memory exhaustion and premature termination of the checker.
Kani~\cite{kani}, a bit-precise BMC for Rust, effectively verifies unsafe Rust code. It can detect memory safety violations such as null pointer dereferences and use-after-free errors, as well as runtime panics from unexpected behaviors like index-out-of-bounds accesses and arithmetic overflows. Kani uses proof harnesses with symbolic inputs to analyze programs. To generate symbolic inputs for a type, the type must implement the \code{kani::Arbitrary} trait, which is already available for most primitive and many standard library types, but user-defined types may require manual implementations. Kani has been successfully applied to verify several real-world Rust projects~\cite{kani-s2n-quic, kani-hifitime, kani-firecracker}. In this paper, we leverage Kani to check the harnesses generated by LLMs.

\noindent\textbf{Large Language Models.}
Large language models (LLMs) have been widely applied to code analysis tasks, including fuzzing~\cite{promptfuzz, ckgfuzz, whitefox}, automated program repair~\cite{llm-apr}, and software testing~\cite{codemosa}. Prompt engineering is a key methodology for interacting with LLMs, employing techniques such as zero-shot prompting~\cite{zero-shot}, few-shot prompting~\cite{few-shot}, Chain-of-Thought~\cite{cot}, and ReAct~\cite{react} to improve output accuracy. In this paper, we exploit the capabilities of LLMs and integrate them into the process of harness generation.

%% file: overview.tex
\section{Methodology}
\subsection{Overview}
\begin{figure}[htb]
    \centering
    \includegraphics[width=0.99\linewidth]{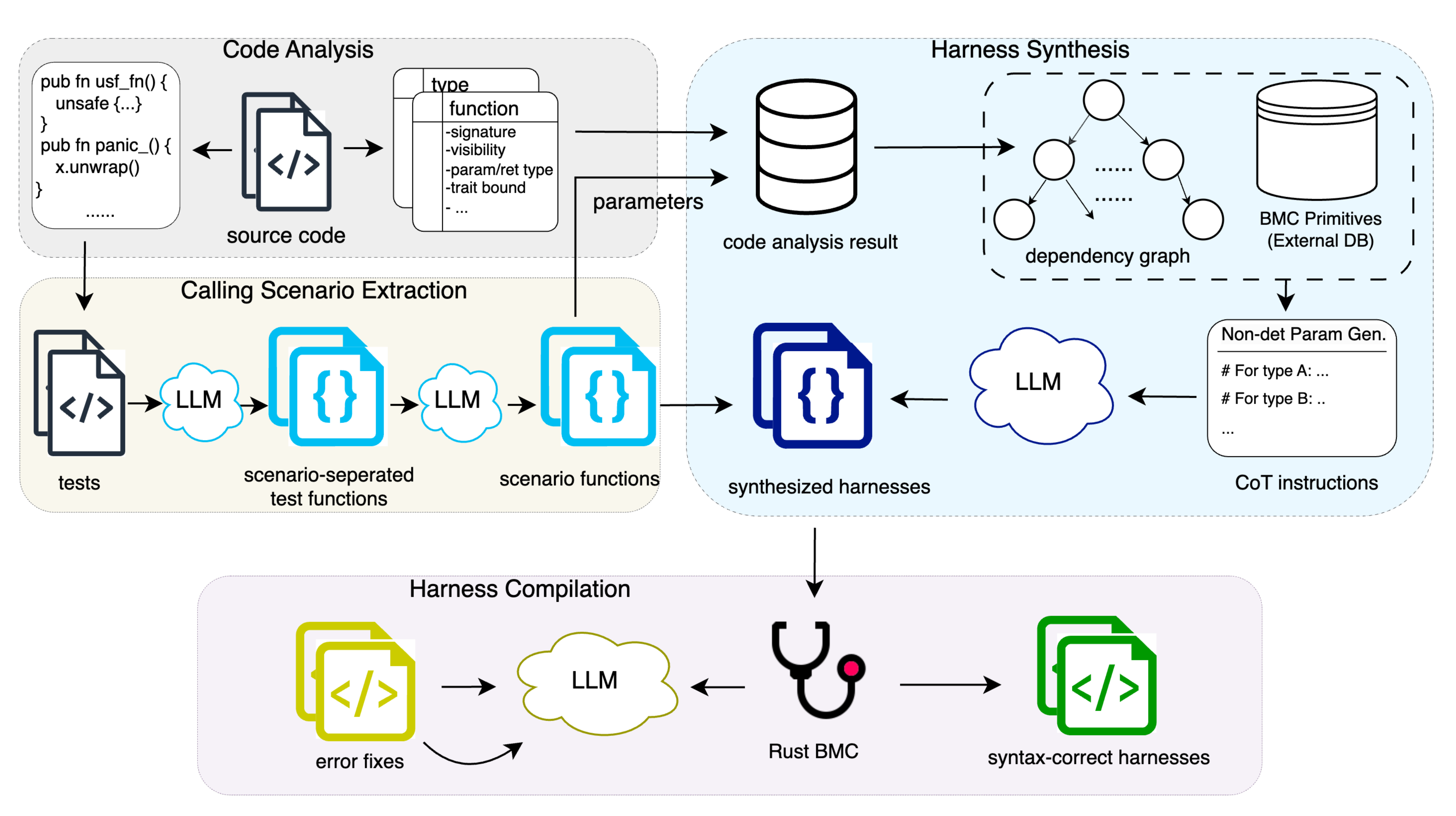}
    \caption{Architecture of \codename.}
    \label{fig:arch}
    \Description{}
    % \vspace{-2ex}
\end{figure}

To facilitate harness generation for eliminating Rust memory safety issues and runtime panics, 
we propose \codename, a method that leverages LLMs to automatically generate harnesses for Rust code. \autoref{fig:arch} outlines our approach, which begins with lightweight code analysis, including parsing the Rust source code for the implementation details of functions and types, and identifying all the APIs containing unsafe and the operations potential causing runtime panics. \codename also locates test cases for those APIs in the codebase. Next, \codename extracts calling scenarios from these test cases and encapsulates each scenario into a separate function, called a \candidate{}. Since these \candidates{} may represent redundant scenarios, \codename instruments the Rust functions and compares their execution traces to eliminate duplicates, finally forming \sketchs{}, with each represents distinct scenario. 

Harness synthesis then begins with generating nondet parameters for each \sketch{}. For each \sketch{}, \codename first checks whether its parameters can be constructed directly using Kani’s primitives. If not, \codename identifies public type constructors for the parameter types and generates nondet data using them. If these constructors require parameters, the process is recursively applied until all required types can be built using Kani’s primitives. During this process, a dependency graph is constructed with each node representing a type constructor (\code{tyf}) and edges pointing to the constructors for \code{tyf}'s parameter types. By traversing this graph, \codename produces Chain-of-Thought (CoT) instructions that include only the essential type construction steps, ensuring they fit within LLMs’ context window. An external knowledge database on common Kani usage further refines these instructions. Finally, a self-contained harness is synthesized by combining the \sketch{}, the nondet parameter generation code, and its invocation.

After synthesis, \codename initiates an iterative harness compilation process. In each iteration, \codename invokes Kani to compile the harness, and LLMs are instructed to fix errors only in the affected code sections. \codename inspects for any fabricated types or functions and generates feedback prompts for targeted fixes. This process continues until all errors are resolved or a predefined iteration limit is reached.

\begin{figure*}[htbp]
    \captionsetup[subfigure]{labelfont={caplabelsmall},justification=centering}
    \centering
    \hfill
    \begin{subfigure}[b]{0.28\textwidth} %  0.30
    \centering
        \begin{ScriptSizePygex}[\verysmall]
            \input{pygtex/test_fn.pygtex} 
        \end{ScriptSizePygex}
        \vspace{-1ex}
        \caption{\scriptsize Test function for \texttt{encode}.}
        \label{fig:run-example-a}
    \end{subfigure}
    \hfill
    \begin{subfigure}[b]{0.35\textwidth} % 0.35
    \centering
        \begin{ScriptSizePygex}[\verysmall]
            \input{pygtex/separate_fn.pygtex}
        \end{ScriptSizePygex}
        \vspace{-1ex}
        \caption{\scriptsize Scenario-separated test functions for \texttt{encode}.}
        \label{fig:run-example-b}
    \end{subfigure}
    % \hfill
    \begin{subfigure}[b]{0.30\textwidth} % 0.32
    \centering
        \begin{ScriptSizePygex}[\verysmall]
            \input{pygtex/scen_fn.pygtex}
        \end{ScriptSizePygex}
        \vspace{-1ex}
        \caption{\scriptsize Scenario function for \texttt{encode}.}
        \label{fig:run-example-c}
    \end{subfigure}
    
    \vspace{2ex} % 在两排之间添加垂直间距

    \hfill
    \begin{subfigure}[b]{0.60\textwidth}
      \hfill
      \centering
        \begin{minipage}{0.45\linewidth}
            \begin{subfigure}{\linewidth}
                \begin{ScriptSizePygex}[\verysmall]
                    \input{pygtex/ty.pygtex}
                \end{ScriptSizePygex}
            \end{subfigure}
            \begin{subfigure}{\linewidth}
                 \includegraphics[width=0.4\linewidth]{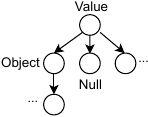}
            \end{subfigure}
        \end{minipage}
        \begin{minipage}{0.50\linewidth}
           \begin{subfigure}{\linewidth}
           \begin{zeropaddingbox}
               \begin{ScriptSizePygex}[\verysmall]
                   \input{pygtex/inst.pygtex}
               \end{ScriptSizePygex}
            \end{zeropaddingbox}
           \end{subfigure}
        \end{minipage}
        % \vspace{-1ex}
        \caption{\scriptsize Type definition, dependency graph and CoT instructions.}
        \label{fig:run-example-d}
    \end{subfigure}
    \hfill
    \begin{subfigure}[b]{0.34\textwidth} % 0.34
        \centering
        \begin{ScriptSizePygex}[\verysmall]
            \input{pygtex/harness.pygtex}
        \end{ScriptSizePygex}
        \vspace{-1ex}
        \caption{\scriptsize Synthesized harness for \texttt{encode}.}
        \label{fig:run-example-e}
    \end{subfigure}
    
    \caption{An example of \codename{}'s workflow.}
    \label{fig:run-example}
    \Description{}
\end{figure*}

\noindent\textbf{A Running Example.}
\autoref{fig:run-example} illustrates \codename's workflow for generating a verification harness for the function \code{encode}. \autoref{fig:run-example-a} presents a test case for the function \code{encode}, from which we extract calling scenarios. Two temporary \candidates{} are generated, representing equivalent scenarios, as seen in \autoref{fig:run-example-b}, and after refinement, we obtain the final \sketch{} \code{scen\_encode\_array} (\autoref{fig:run-example-c}). This function takes a \code{Value} parameter whose definition, as well as that of its dependent type \code{Object}, is shown in \autoref{fig:run-example-d}. To generate nondet arguments, we first construct a dependency graph and then generate a set of Chain-of-Thought instructions based on the graph for the LLM, instructing it to generate a nondet \code{Object} before constructing the \code{Value} (see \autoref{fig:run-example-d}). The final synthesized harness (\autoref{fig:run-example-e}) includes the scenario function, the Kani harness function, and the functions required for constructing nondet data. Finally, the harness is compiled and iteratively refined until a syntactically correct harness for the function \code{encode} is obtained.

\subsection{Code Analysis}
As the first step, we perform a lightweight static analysis of the entire Rust codebase to identify those functions that require verification harnesses. Specifically, we traverse the Rust's MIR to find all functions that either contain unsafe blocks or can trigger a runtime panic. We focus solely on functions implemented within the Rust codebase, excluding those from the Rust standard library and third-party dependencies. After identifying these target functions, we locate all test cases that target them. These test cases serve as the sources for extracting calling scenarios. 

Harness synthesis requires creating nondet inputs, so we must locate, for every parameter type, the public constructors that can produce values of that type. To this end, we scan \code{impl Ty} blocks for public methods whose return type is \code{Self}, \code{Ty}, \code{Result<Ty>}, or \code{Option<Ty>}. When such methods include documentation examples or doctests in their comments, we extract those examples as well, since they often demonstrate valid usage patterns.
We also gather the full definitions and visibility attributes of all types and traits declared in the code, recording which traits each type implements and which types implement each trait. This information enhances the accuracy and reliability of type dependency analysis, thereby facilitating effective harness synthesis.

\subsection{Calling Scenario Extraction}
\label{sec:sketch_func_gen}
Our scenario extraction approach builds on two key observations. First, test functions typically use assert statements to exercise target APIs, with each assert representing a distinct invocation scenario. Second, the literal constants in those tests reveal valid API inputs and should be treated symbolically during verification to cover a broader input space. 
Based on these insights, we first extract the statement sequence leading up to each assert and encapsulate it into a separate function, called a \candidate{}. Next, we promote the constants within the sequence to function parameters, resulting in a new function, which we denote as a \sketch{}. Finally, to ensure that the extracted scenario preserves the original behavior, we replay each \sketch{} with the same constants as in the \candidate{}, compare its execution trace to that of the original. If they match, we consider that the invocation scenario has been accurately preserved.

Traditional program analysis tools, such as custom LLVM or MIR passes, could automate parts of this process, but they often struggle with Rust’s generics, closures, and higher-order functions and frequently break across different Rustc or LLVM versions. By contrast, LLMs handle these challenges smoothly when guided by well-designed prompts, enabling a lightweight, robust implementation. Therefore, we integrate LLMs into this process. 

\noindent\textbf{{Scenario-separated Test Functions.}}
In this stage, we extract individual calling scenarios by encapsulating each assert and its dependent code into a separate function, namely, a \candidate{}. 
As shown in \autoref{fig:prompt_assert_split}, the prompt directs the LLM to identify all assert statements, perform backward dataflow analysis on the operands to locate all dependent variables and statements, and then organize these statements into separate functions with specified names.

\autoref{fig:run-example-b} shows an example of two extraced \candidates{} (\code{encode\_with\_array\_macro} and \code{encode\_with\_push\_null}) representing identical invocation scenarios, which should be represented by \code{scen\_encode\_array}. 
To prevent redundant work in downstream harness generation, we deduplicate \candidates{} so that each distinct invocation scenario is represented only once. This is achieved by instrumenting function entry points in the Rust code to trace execution, running all \candidates{}, and comparing their traces to identify and remove redundant functions.

\begin{figure}[htb]
    \centering
    \resizebox{\linewidth}{!}{%
        \begin{zeropaddingbox}
            \begin{ScriptSizePygex}
                \input{pygtex/assert_split.pygtex}
            \end{ScriptSizePygex}
        \end{zeropaddingbox}
    }
    \caption{Scenario-separated test function generation prompt.}
    \label{fig:prompt_assert_split}
    \Description{}
    % \vspace{-2ex}
\end{figure}

\noindent\textbf{{Scenario Function Generation}.}
A scenario function represents a distinct invocation scenario of a target API, with its parameters capturing all possible input values for that scenario. 
We generate \sketchs{} by refactoring \candidates{} through a process of assert removal, constant promotion, and parameter refinement. This workflow leverages prompt chaining~\cite{prompt-chaining} (\autoref{fig:prompt-chain}), which is particularly effective for complex tasks that might overwhelm LLMs if addressed with a very detailed prompt.

\begin{figure*}[htbp]
    \begin{subfigure}[b]{0.38\linewidth}
        \begin{zeropaddingbox}
            \begin{ScriptSizePygex}
                \input{pygtex/assert_remove.pygtex}
            \end{ScriptSizePygex}
        \end{zeropaddingbox}
        \caption{\footnotesize Assert removal.}
        \label{fig:prompt_assert_decons}
    \end{subfigure}
    \hfill
     \begin{subfigure}[b]{0.32\linewidth}
        \begin{zeropaddingbox}
            \begin{ScriptSizePygex}
                \input{pygtex/gen_sketch.pygtex}
            \end{ScriptSizePygex}
        \end{zeropaddingbox}
        \caption{\footnotesize Constant promotion.}
        \label{fig:prompt_gen_sketch}
    \end{subfigure}
    \hfill
     \begin{subfigure}[b]{0.28\linewidth}
        \begin{zeropaddingbox}
            \begin{ScriptSizePygex}
                \input{pygtex/remove_unused.pygtex}
            \end{ScriptSizePygex}
        \end{zeropaddingbox}
        \caption{\footnotesize Parameter refinement.}
        \label{fig:prompt_remove_unused}
    \end{subfigure}
    \vspace{-2ex}
    \caption{Prompt chaining for \sketch{} generation.}
    \label{fig:prompt-chain}
    \Description{}
\end{figure*}

First, we remove the assert statement from \candidates{}, as our focus is on memory safety and runtime panic issues rather than functional correctness. This is achieved by replacing asserted expressions with separate variables and returning a variable that represents a comprehensive execution context, as depicted in~\autoref{fig:prompt_assert_decons}.

Next, we promote the constants in the scenario-separated test function to parameters of the scenario function while ensuring that the function returns specified variables. The LLM is prompted to identify all actively used constants that are assigned to variables and used in subsequent statements, and promote them as parameters without altering the function’s semantics, as shown in \autoref{fig:prompt_gen_sketch}.

Finally, we refine the scenario function's parameters by removing any that are not essential for the invocation scenario. In the harness synthesis stage, each parameter is assigned nondet values, and superfluous parameters cause unnecessary LLM generations. We observed that removing unused variables can help eliminate unused parameters. As Rustc flags unused variables during compilation, we iteratively instruct the LLM with the prompt shown in \autoref{fig:prompt_remove_unused} to remove unused parameters by leveraging Rustc warnings about unused variables. 

\noindent\textbf{{Preservation of Calling Scenarios.}}
To validate that each scenario function preserves its original calling context, we automatically generate test cases for it by reusing the literal constants from its corresponding scenario-separated test function. We invoke the scenario function with the same constants and apply identical assertions, then compare the execution traces of both the scenario function and the original scenario-separated test function. If the traces match, the scenario function is considered to have preserved the invocation scenario. Otherwise, it is excluded from further harness generation.
The test suites may also contain flaky tests whose outcomes vary across different runs. To address this, we run each scenario function multiple times and aggregate all observed traces into a set, ensuring we capture as many execution paths as possible and avoid transient mismatches. 
Test case generation for \sketchs{} is driven by the prompt shown in \autoref{fig:prompt_sketch_test}, which specifies the relationship between the \sketch{} and its original test. With prior instrumentation in place, we execute both versions, collect their traces, and perform the comparisons.

\begin{figure}[htb]
    \begin{zeropaddingbox}
       \begin{subfigure}{0.55\columnwidth}
            \begin{ScriptSizePygex}
                \input{pygtex/sketch_test_p1.pygtex}
            \end{ScriptSizePygex}
       \end{subfigure}
       \hfill
       \begin{subfigure}{0.40\columnwidth}
             \begin{ScriptSizePygex}
                \input{pygtex/sketch_test_p2.pygtex}
            \end{ScriptSizePygex}
       \end{subfigure}
   \end{zeropaddingbox}
    \caption{Prompt of generating tests for \sketchs{}.}
    \label{fig:prompt_sketch_test}
    \Description{}
\end{figure}

\subsection{Harness Synthesis}\label{sec:harn-synthsis}
We leverage the prompt shown in \autoref{fig:gen_harn_for_sketch_fn} to synthesize harnesses. The harness is constructed in a separate Rust module, with Kani harness functions attributed with \code{\#[kani::proof]}, and necessary functions for nondet type construction. 

The most challenging aspect, highlighted in the figure, is the construction for nondet parameters of the \sketch{}.
Firstly, Kani provides functions for constructing nondet types, but these are limited to primitive Rust types. For complex types, although Kani offers the \code{Arbitrary} trait to allow users to manually construct nondet data, it requires adding this to all dependent data types. This requires extensive manual modifications to the codebase, which is error-prone. Additionally, the construction of nondet types must account for Rust's type conversions. For example, some parameters may be traits, requiring the identification of all concrete types that implement the trait.

To overcome these challenges, we adopt the following strategy for constructing nondet types. If Rust basic types can be built with Kani's primitives, we directly utilize them for construction. Otherwise, we use the type's public constructor to create it. The same approach is applied recursively to the parameters of constructor functions. Through this process, we can incrementally build nondet data for all required types. This strategy can be represented by constructing a dependency graph, where each of the other nodes represents a type constructor, and its edges point to the constructors of the parameters required by that type constructor. Finally, the steps to build parameters of \sketch{} can be determined by performing a topological traversal of the dependency graph. 

\begin{figure}[htb]
    \begin{zeropaddingbox}
        \begin{subfigure}{0.42\linewidth}
            \begin{ScriptSizePygex}
                \input{pygtex/harn_syn_p1.pygtex}
            \end{ScriptSizePygex}
        \end{subfigure}
        \hfill
        \begin{subfigure}{0.56\linewidth}
            \begin{ScriptSizePygex}
                \input{pygtex/harn_syn_p2.pygtex}
            \end{ScriptSizePygex}
        \end{subfigure}
    \end{zeropaddingbox}
    \caption{Harness synthesis prompt with detailed nondet type construction steps (highlighted). }
    \label{fig:gen_harn_for_sketch_fn}
    \Description{}
\end{figure}

\noindent{\textbf{Dependency Graph.}} 
\autoref{algo:type_dep_ana} describes the process of building the dependency graph, denoted as \code{G}. In this algorithm, 
\code{cnstr\_fn} represents a type constructor and \code{fn\_par\_tys} represents the constructor's parameters. \code{db} stores the results of code analysis, and \code{llm} refers to the LLM utilized to generate the necessary code for constructing the nondet data.
The algorithm begins by fetching example usages of the constructor from \code{db}, the codebase's documentation tests, providing high-quality references for the LLM (line 3-4). 
It then checks whether the parameters of the constructor function are traits. If so, it queries \code{db} to identify all concrete types that implement the trait (line 7-9). 
The LLM is then prompted to generate nondet data for these concrete types using prompts as illustrated in \autoref{fig:ask_nondet_gen_first} (line 11). 
For a concrete type (\code{par\_ty}), its definition is provided to the LLM, and if the LLM successfully generates the construction code, this code is wrapped into a new function (serving as the ``nondet constructor" for \code{par\_ty}) and added as a node to the dependency graph (line 14-16). If generation fails, the algorithm retrieves \code{par\_ty}'s constructors from \code{db} (via \code{db.get\_constructors\_for}), fetches the parameters of each constructor, links each constructor from the current node, and recursively invokes itself (line 19-25). Note that definitions of \code{pub enum} and \code{pub struct} with all public fields are also treated as their constructors, as they can be directly instantiated through their fields, which serve as the parameters.
In the initial invocation of this algorithm, the parameters \code{cnstr\_fn} and \code{fn\_par\_tys} correspond to \sketch{} and its parameters, making the \sketch{} the root node of the graph.

\begin{figure}[htb]
\begin{zeropaddingbox}
     \begin{subfigure}{0.49\columnwidth}
        \begin{ScriptSizePygex}
            \input{pygtex/rec-kani-non-det-pub-fn-ask.pygtex}
        \end{ScriptSizePygex}
    \end{subfigure}
    \hfill
    \begin{subfigure}{0.48\columnwidth}
        \begin{ScriptSizePygex}
            \input{pygtex/kani-primitives.pygtex}
        \end{ScriptSizePygex}
    \end{subfigure}
\end{zeropaddingbox}
    \caption{The left is the nondet type generation prompt used in \autoref{algo:type_dep_ana}, with the highlighted representing embedded external knowledge shown on the right.}
    \label{fig:ask_nondet_gen_first}
    \Description{}
\end{figure}

Additionally, the external knowledge embedded in the prompt in \autoref{fig:ask_nondet_gen_first} includes specific, pre-implemented and syntactically correct code examples that demonstrate how to use Kani to generate nondet data for common scenarios, including primitive types, enums, and structs. These examples enable the LLM to reference or directly adopt correct code during generation, improving  compilation success and reducing subsequent harness compilation overhead.

\SetAlgoBlockMarkers{begin}{end} 
% setup tcc style
\renewcommand{\tcc}[1]{%
  \par\vskip.1ex%
  \textit{/*\ #1\ */}%
  \par\vskip.1ex%
}
\newcommand{\MyComment}[1]{\tcc{\textcolor{purple}{#1}}}

\newcommand{\var}[1]{\textit{#1}} % variable stype
\newcommand{\func}[1]{\texttt{#1}} % func style

\SetKwProg{Fn}{Function}{:}{end} 
\SetKwFunction{GenDepGraph}{GEN\_DEP\_GRAPH}
\SetKwFunction{NoCircle}{\func{no\_circle}}
\SetKwFunction{AddNode}{\func{add\_node}}
\SetKwFunction{AddEdge}{\func{add\_edge}}
\SetKwFunction{GetUsages}{get\_usages}
\SetKwFunction{GetImplTrait}{\func{get\_impl\_trait}}
\SetKwFunction{GenNondetFor}{\func{gen\_nondet\_for}}
\SetKwFunction{WrapFn}{\func{wrap\_fn}}
\SetKwFunction{GetConstructorsFor}{\func{get\_constructors\_for}}
\SetKwFunction{GetParamTypes}{\func{get\_param\_types}}
\SetKwFor{ForEach}{for}{do}{end for}

\SetInd{0.3em}{0.4em} % {嵌套缩进}{块内缩进}
\SetAlgoInsideSkip{small} % 设置内部间距为小
\SetNlSkip{0.3em} % 设置行间距

\begin{algorithm}[htb]
% \SetAlgoLined
\caption{Generate Dependency Graph}
\label{algo:type_dep_ana}

\Fn{\GenDepGraph{\var{G}, \var{llm}, \var{db}, \var{cnstr\_fn}, \var{fn\_par\_tys}}}{
    \MyComment{Get example usages for \var{cnstr\_fn}} 
    \var{examples} $\gets$ \var{db}.\GetUsages{\var{cnstr\_fn}} \\
    \var{G}.\AddNode{\var{cnstr\_fn}, \var{examples}} %\MyComment{Fill the node with examples}

    \ForEach{var{par\_ty} $\in$ \var{fn\_par\_tys}}{
        \MyComment{Get types that impl \var{par\_ty}}
        \var{ty\_impl} $\gets$ \var{db}.\GetImplTrait{\var{par\_ty}, \var{cnstr\_fn}} %\MyComment{Get types that impl \var{par\_ty}}
        
        \If{\var{ty\_impl} is not empty}{
            \var{par\_ty} $\gets$ \var{ty\_impl}
        }
        \MyComment{Ask LLM to gen nondet \var{par\_ty}}
        \var{ans} $\gets$ \var{llm}.\GenNondetFor{\var{par\_ty}} %\MyComment{Ask LLM to gen nondet \var{par\_ty}}

        \eIf{\var{ans} $\neq$ 'NULL'} {
            \MyComment{If LLM generates nondet \var{par\_ty}, wrap it as its nondet constructor}
            % \MyComment{Wrap as nondet constructor of \var{par\_ty}}
            \var{nondet\_fn} $\gets$ \WrapFn{\var{par\_ty}, \var{ans}}
            
            \If{\NoCircle{\var{G}, \var{nondet\_fn}, \var{cnstr\_fn}}}{
                \var{G}.\AddEdge{\var{cnstr\_fn}, \var{nondet\_fn}}
            }
        }{ 
            \MyComment{Otherwise, look for \var{par\_ty}'s constructors}
            \ForEach{\var{par\_cnstr\_fn} $\in$ \var{db}.\GetConstructorsFor{\var{par\_ty}}}{ 
                \If{\NoCircle{\var{G}, \var{cnstr\_fn}, \var{par\_cnstr\_fn}}}{ %\MyComment{no circular dependency}
                    \MyComment{\var{cnstr\_fn} points to its param's constructor \var{par\_cnstr\_fn}}
                    \var{G}.\AddEdge{\var{cnstr\_fn}, \var{par\_cnstr\_fn}} \\ 
                    \var{par\_fn\_tys} $\gets$ \var{db}.\GetParamTypes{\var{par\_cnstr\_fn}} 
                    
                    \MyComment{Recursive call}
                    \GenDepGraph{\var{G}, \var{llm}, \var{db}, \var{par\_cnstr\_fn}, \var{par\_fn\_tys}} %\MyComment{Recursive call}
                }
            }
        }
    }
}
\end{algorithm}

\noindent{\textbf{CoT Instructions for Synthesis.}}
We then perform a topological traversal of the dependency graph to generate CoT instructions that guide LLMs in incrementally producing nondet arguments for the \sketch{}. These instructions, formatted in Markdown, allocate one section per graph node. For leaf nodes (those with no outgoing edges), each section provides code snippets or examples to construct the corresponding nondet types. For non-leaf nodes, which include public constructor functions, \code{pub struct} (with all public fields), or \code{pub enum}, the sections describe how to create nondet parameters or fields, and indicate which sections should be referenced in the construction process. \autoref{fig:CoT_inst_templ} illustrates the structure of the output CoT instructions. 

\begin{figure}[htb]
    \begin{zeropaddingbox}
        \begin{subfigure}{0.50\textwidth}
             \begin{ScriptSizePygex}
                 \input{pygtex/cot-inst-templ-p1.pygtex}
             \end{ScriptSizePygex}
        \end{subfigure}
        \hfill
         \begin{subfigure}{0.48\textwidth}
             \begin{ScriptSizePygex}
                 \input{pygtex/cot-inst-templ-p2.pygtex}
             \end{ScriptSizePygex}
        \end{subfigure}
    \end{zeropaddingbox}
    \caption{Structure of CoT instructions for LLMs to generate nondet parameters.}
    \label{fig:CoT_inst_templ}
    \Description{}
\end{figure}

% \begin{figure}[htb]
%     \begin{zeropaddingbox}
%         \begin{subfigure}{0.50\textwidth}
%              \begin{ScriptSizePygex}
%                  \input{pygtex/cot-inst-templ-p1.pygtex}
%              \end{ScriptSizePygex}
%         \end{subfigure}
%       \begin{subfigure}{0.48\textwidth}
%              \begin{ScriptSizePygex}
%                  \input{pygtex/cot-inst-templ-p2.pygtex}
%              \end{ScriptSizePygex}
%         \end{subfigure}
%     \end{zeropaddingbox}
%     \caption{Structure of CoT instructions for LLMs to generate nondet parameters.}
%     \label{fig:CoT_inst_templ}
%     \Description{}
% \end{figure}

\subsection{Harness Compilation}\label{sec:harn-compilation}
After synthesizing harnesses, we invoke Kani to compile them and use LLMs to resolve compilation errors. To ensure successful compilation, \codename makes the code self-contained by including the \sketch{} and the nondet data generation functions from the external knowledge database in the synthesized code.
\codename invokes Kani to compile the entire code. If Kani fails to compile, \codename constructs a feedback prompt containing the error messages and instructs the LLM to fix the issues. This process is repeated until the LLM successfully resolves all compilation errors or the predefined iteration limit is reached.  

\begin{figure}[htb]
    \begin{subfigure}{0.9\columnwidth}
        \begin{zeropaddingbox}
            \begin{ScriptSizePygex}
                \input{pygtex/err_fix_simple.pygtex}
            \end{ScriptSizePygex}
        \end{zeropaddingbox}
        \caption{\footnotesize Simple error-fix prompt (\texttt{simple-err-fix}).}
        \label{fig:err_feedback_simple}
    \end{subfigure}
    \hfill
    \begin{subfigure}{0.9\columnwidth}
        \begin{zeropaddingbox}
            \begin{ScriptSizePygex}
                \input{pygtex/err_fix.pygtex}
            \end{ScriptSizePygex}
        \end{zeropaddingbox}
        \caption{\footnotesize Error-fix prompt (\texttt{err-fix}).}
        \label{fig:err_feedback}
    \end{subfigure}
    \caption{Different prompts designed for error fixing.}
    \Description{}
\end{figure}

\autoref{fig:err_feedback_simple} showcases a simple error-fix prompt, which is not effective (as tested in \autoref{sec:ablation-study}).
As the \sketch{} and the external knowledge functions are already syntactically correct, they should remain unchanged during the error-fixing process. Therefore, we designed the prompt shown in \autoref{fig:err_feedback} to guide the LLM in error correction. The LLM is instructed to generate the responses in three parts: \sketch{}, which is guaranteed to be syntactically correct as ensured by the generated test cases (\autoref{sec:sketch_func_gen}), pre-implemented and syntactically correct functions from the external knowledge database, and the necessary error fixes. 
It is crucial that the first two parts remain unchanged throughout the error-fixing process, as explicitly emphasized in the prompt. Our experiments have shown that omitting these components, for instance, using the prompt shown in \autoref{fig:err_feedback_simple}, can increase the number of generation attempts. This is because LLMs may become distracted from the core task of error correction or introduce additional errors. 
Furthermore, due to potential hallucinations, the LLM may fabricate types to ensure the code compiles with Kani. To mitigate this, we introduce a lightweight AST checker inspecting whether a type already defined in the codebase is redefined in the LLM-generated code. If such duplication is detected, we notify the LLM of the error and instruct it to generate new code without introducing any new types.

%% file: pygtex/assert_split.pygtex
\begin{Verbatim}[commandchars=\\\{\}]
\PYG{l+s+sb}{```rust}
\PYG{o}{\PYGZlt{}}\PYG{n}{test\PYGZus{}fn\PYGZus{}code\PYGZus{}mutilple\PYGZus{}asserts}\PYG{o}{\PYGZgt{}}
\PYG{l+s+sb}{```}
Separate the function into multiple test functions, each with a single 
\PYG{l+s+sb}{`assert`}.
\PYG{g+gh}{\PYGZsh{} Instructions:}
1. Perform a backward use\PYGZhy{}def analysis on each asserted expression and 
identify all dependent statements.
2. Extract the identified dependent statements and the corresponding \PYG{l+s+sb}{`assert`} 
to form a new test function.
3. Name the new function as \PYG{l+s+sb}{`scen\PYGZus{}separated\PYGZus{}test\PYGZus{}\PYGZlt{}X\PYGZgt{}`}.\PYG{l+s+sb}{`\PYGZlt{}X\PYGZgt{}`} starts from 1.
\PYG{g+gh}{\PYGZsh{} Output:}
\PYG{l+s+sb}{```rust}
\PYG{c+cp}{\PYGZsh{}[test]}\PYG{+w}{ }\PYG{k}{fn} \PYG{n+nf}{scen\PYGZus{}separated\PYGZus{}test\PYGZus{}}\PYG{o}{\PYGZlt{}}\PYG{n}{X}\PYG{o}{\PYGZgt{}}\PYG{p}{()}\PYG{+w}{ }\PYG{p}{\PYGZob{}}
\PYG{+w}{ }\PYG{c+c1}{// statements that have a data dependency on the asserted expression}
\PYG{p}{\PYGZcb{}}
\PYG{l+s+sb}{```}
\end{Verbatim}

%% file: impl.tex
\section{Implementation}
We implemented \codename in about 3,000 lines of Python code and 260 lines of Rust code.

\noindent{\textbf{Preprocess.}}
We apply function tracing to refine redundant calling scenarios. Using the logfn Rust crate~\cite{logfn}, we instrument function entries by inserting the attribute \code{\#[logfn::logfn(Pre,Debug,{fn})]} before each function. This logs function execution when the environment variable \code{RUST\_LOG} is set to \code{debug}. Function names are then extracted from log lines containing \code{DEBUG} to form the traces. This instrumentation is performed prior to our workflow.

\noindent{\textbf{Code Analysis.}}
We analyze Rust’s MIR to identify unsafe code and potential runtime panics. Runtime panics are detected by matching \code{Option/Result::unwrap} calls and compiler-inserted assertions at the end of MIR basic blocks. Unsafe code is identified using each function’s MIR safety property \code{(MIR.source\_scopes.local\_data.s\\afety)}, which marks functions containing or declared as unsafe. All analyses are implemented as a Rustc plugin.

To detect fabricated types or functions during harness compilation, we use tree-sitter~\cite{tree-sitter} to parse the AST of LLM-generated harness code. By comparing parsed definitions with those in the original codebase, we can identify any fabricated code.

\noindent{\textbf{Rust Compilers.}}
Our workflow employs two Rust compilers: Rustc for eliminating unused parameters during the \sketch{} generation process, and Kani-0.63.0~\cite{kani} to compile the LLM-generated harnesses. Kani provides error messages that are fed back to the LLM for iterative error correction.

\noindent{\textbf{LLM Settings.}}
We implement \codename atop OpenAI's GPT-4.1 API~\cite{gpt-4.1-2025-04-14} (gpt-4.1-2025-04-14), with the temperature fixed at 0 and all other parameters set to default. In all iterative interactions with Rust compilers, we limit the number of iterations to 10.

%% file: eva.tex
\section{Evaluation}
Our evaluation aims to address the following research questions.
\begin{itemize}[leftmargin=1em]
    \item RQ1: How does \codename perform harness generation for real-world Rust codebases?
    \item RQ2: How do the key components of \codename contribute to the effectiveness?
    \item RQ3: How does \codename compare against existing harness generation methods?
    \item RQ4: How does \codename perform when applied with different LLMs?
\end{itemize}
We evaluated these questions using GPT-4.1 (gpt-4.1-2025-04-14). For RQ4, we also test with claude-sonnet-4-20250514~\cite{claude-4}, DeepSeek-v3-0324~\cite{ds-v3} and DeepSeek-R1-0528~\cite{ds-r1} using the same parameter settings as GPT-4.1. 

\subsection{Settings}
\noindent\textbf{Dataset.} We used two datasets to evaluate \codename:
\begin{itemize}[leftmargin=0pt, label=\textbullet, itemsep=0pt]
    \item \textbf{Real-world Rust libraries ($D_{rwd}$)}. We selected \numofcrates{} Rust libraries from various categories on crates.io~\cite{crates-io}, comprising a total of \teststotalnumber{} tests and \testslineofcode{} lines of code. Details are presented in \autoref{tab:dataset}. This dataset was used to assess the overall effectiveness of \codename.
    \item \textbf{Scenario functions dataset ($D_{scen}$)}. \numofsketchs{} \sketchs{} were extracted during the harness generation on ${D_{rwd}}$. These \sketchs{} were used to evaluate the individual contributions of harness synthesis and compilation.
\end{itemize}

\noindent\textbf{Platform.} The evaluation was conducted using an Intel(R) Xeon(R) CPU E5-2673 v4 @ 2.30GHz with 80 cores and 256GB RAM, and a 1TB hard drive, running Ubuntu 20.04.6 LTS.

\begin{table}[htb]
    \centering
    \arrayrulecolor{gray!100}
    {\footnotesize
    \begin{tabular}{c|c|c|c|c}
        \toprule
        \textbf{Library} & \textbf{Category} & \textbf{\#Test Files} & \textbf{\#Test Funcs} & \textbf{LoC} \\
        \midrule
        pdf-rs & pdf & 12 & 34 & 521 \\  % pdf
        % \hline
        tar-rs & tar / encoding & 4 & 100 & 2387 \\ % tar / encoding
        % \hline 
        jpeg-decoder & image / decoder & 9 & 21 & 447 \\ % image / decoder / jpeg
        % \hline 
        tempfile & filesystem & 5 & 52 & 637 \\ % filesystem
        % \hline 
        jzon-rs & json / serialization  & 9 & 192 & 1429 \\ 
        % \hline 
        image-webp & encoding / decoding & 4 & 18 & 291 \\ % webp encoding / decoding
        lexical-util & numeric conversion & 9 & 26 & 442 \\
        prost-types & prost definitions & 5 & 15 & 151 \\
        p256 & elliptic curve & 6 & 36 & 346 \\
        \midrule 
        \textbf{Total} & / & 63 & \teststotalnumber{} & \testslineofcode{} \\
        \bottomrule
    \end{tabular}
    }
    \caption{Details of Rust libraries in $D_{rwd}$, including each name, category, number of test files and functions, and total lines of test functions.}
    \label{tab:dataset}
    % \vspace{-5ex}
\end{table}

\subsection{RQ1: Effectiveness} 
\noindent\textbf{Calling Scenarios Extraction}.
\autoref{tab:eff-call-scen} summarizes the calling scenarios extracted by \codename from the libaries in $D_{rwd}$. The \code{\#Scenarios} column indicates the number of calling scenarios extracted from the tests. The \code{\#Prsv.} column shows the percentage of scenarios that accurately preserve the original calling scenarios within the test functions. In contrast, the \code{\#Non-Prsv.} column reflects the percentage of scenarios that differ from the original test functions. 
As detailed in \autoref{sec:sketch_func_gen}, we instrument the functions within these libraries, generate tests for the extracted calling scenarios, execute them to obtain function traces, and compare these traces with those from the original test functions. A scenario is considered correctly preserved if the traces match.

Our results show that, across the \numofcrates{} libraries, \scenpresvrate{} of calling scenarios were preserved on average, with \code{image-webp}, \code{lexical-util}, and \code{jpeg-decoder} achieving full preservation. This demonstrates that our approach effectively extracts calling scenarios, laying a solid foundation for harness generation.

\begin{table}[htb]
    \centering
    \arrayrulecolor{gray!100}
    {\footnotesize
    \begin{tabular}{c|c|c|c|c}
        \toprule
        \textbf{Library} & \textbf{\#Scenarios} & \textbf{\#Prsv.} & \textbf{\#Non-Prsv.} & \textbf{Prsv. Rate} \\
        \midrule
        % pdf-0.9.0
        pdf-rs & 27 & 25 & 2 & 92.59\% \\  
        % \hline
        tar-rs & 82 & 80 & 2 & 97.56\% \\
        % \hline 
        jpeg-decoder & 7 & 7 & 0 & 100\%\\ 
        % \hline 
        tempfile & 36 & 35 & 1 & 97.22\% \\
        % \hline
        % jzon-0.12.5
        jzon-rs & 80 & 65 & 15 & 81.25\% \\
        % \hline
        % image-webp-0.1.3
        image-webp & 9 & 9 & 0 & 100\% \\ 
        lexical-util & 27 & 27 & 0 & 100\% \\
        prost-types & 6 & 5 & 1 & 83.33\% \\
        p256 & 19 & 19 & 0 & 100\% \\
        \midrule 
        \textbf{Average} & 32 & 30 & 2 & 94.66\% \\
        \bottomrule
    \end{tabular}
    }
    \caption{Effectiveness of calling scenarios extraction.}
    \label{tab:eff-call-scen}
\end{table}

\noindent\textbf{Harness Generation}.
\autoref{tab:eff-harn-gen} presents the results of generated harnesses, with the \code{FULL} column showing the results obtained through the complete workflow. The \code{\#Harn.} column indicates the number of harnesses generated by \codename for each Rust library. As described in \autoref{sec:harn-compilation}, during the harness compilation phase, the LLM interacts iteratively with Kani to refine the harnesses, with the number of interactions denoted as \code{@Pass}.
The \code{@Pass=0} column shows the proportion of harnesses that compiled correctly immediately after the synthesis stage. The columns \code{@Pass<=3,5,10} show the proportions of harnesses that successfully compiled within 3, 5, and 10 interactions with the LLM, respectively, relative to the total number of harnesses (\code{\#Harn.}). Finally, the \code{\#Generations} column presents the total number of harness generation attempts when the interaction limit is set to 10.

\newcommand{\dn}{$\downarrow$}
\newcommand{\up}{$\uparrow$}
\newcommand{\lr}{$\leftrightarrow$}
\begin{table*}[htb]
    \centering
    \arrayrulecolor{gray!100}
    {\footnotesize
    \begin{tabular}{c|c|c|c|c|c|c|c|c|c|c|c|c|c|c|c|c}
        \toprule
        \multirow{2}{*}{\textbf{Library}} & \multirow{2}{*}{\textbf{\#Harn.}} & \multicolumn{3}{c|}{\textbf{@Pass=0}} & \multicolumn{3}{c|}{\textbf{@Pass$\leq$3}} & \multicolumn{3}{c|}{\textbf{@Pass$\leq$5}} & \multicolumn{3}{c|}{\textbf{@Pass$\leq$10}} & \multicolumn{3}{c}{\textbf{\#Generations}} \\
        \cline{3-17}
        \noalign{\vskip 0.5ex}
        ~ & ~ & \textbf{FULL} & \textbf{SNG} & \textbf{SEF} & \textbf{FULL} & \textbf{SNG} & \textbf{SEF} & \textbf{FULL} & \textbf{SNG} & \textbf{SEF} & \textbf{FULL} & \textbf{SNG} & \textbf{SEF} & \textbf{FULL} & \textbf{SNG} & \textbf{SEF} \\
        \midrule
        pdf-rs & 27 & 63\% & \dn 37\% & \dn 19\% & 93\% & \dn 12\% & \dn 4\% & 97\% & \up 3\% & \dn 1\% & 100\% & \lr & \dn 4\% & 46 & \up37\% & \up22\% \\
        % \hline
        tar-rs & 82 & 70\% & \dn 11\% & \dn 21\% & 98\% & \dn 4\% & \dn 5\% & 100\% & \lr & \lr & 100\% &  \lr & \lr & 137 & \up13\% & \up20\% \\ 
        % \hline 
        jpeg-decoder & 7 & 86\% &\dn 29\% & \lr & 100\% & \lr & \lr & 100\% & \lr & \lr & 100\% & \lr & \lr & 11 & \up22\% & \lr \\
        % \hline
        tempfile & 36 & 81\% & \dn 28\% & \lr & 100\% & \dn 3\% & \lr & 100\% & \lr & \lr & 100\% & \lr & \lr & 44 & \up 34\% & \lr \\
        % \hline
        jzon-rs & 80 & 83\% & \dn 30\% & \dn 37\% & 95\% & \lr & \dn 2\% & 96\% & \dn 4\% & \dn 1\% & 100\% & \dn 2\% & \lr & 107 & \up34\% & \up 38\% \\
        % \hline
        image-webp & 9 & 100\% & \dn 22\% & \lr & 100\% & \lr & \lr & 100\% & \lr & \lr & 100\% & \lr & \lr & 9 & \up 22\% & \lr \\
        % \hline 
        lexical-util & 27 & 63\% & \dn 9\% & \dn 2\% & 100\% & \dn 11\% & \lr & 100\% & \dn 11\% & \lr & 100\% & \dn 7\% & \lr & 37 & \up 54\% & \lr \\
        % \hline
        prost-types & 6 & 100\% & \dn 33\% & \dn 67\% & 100\% & \lr & \dn 33\% & 100\% & \lr & \lr & 100\% & \lr & \lr & 6 & \up 50\% & \up 133\% \\ 
        % \hline
        p256 & 20 & 75\% & \dn 25\% & \dn 25\% & 90\% & \lr & \dn 25\% & 100\% & \dn 5\% & \dn 5\% & 100\% & \lr & \lr & 31 & \up 32\% & \up 61\% \\
        \hline
        \textbf{Average} & 32 & 80.6\% & \dn 23.1\% & \dn 18.7\% & 97.3\% & \dn 3.3\% & \dn 7.7\% & 97.8\% & \dn 1.9\% & \dn 0.8\% & 100\% & \dn 1.0\% & \dn 0.4\% & 46 & \up 28.3\% & \up 23.9\% \\
        \bottomrule
    \end{tabular}
    }
    \caption{Comparison of harness generation effectiveness using the complete workflow (\textbf{FULL}), the \code{simple-nondet-gen} prompt (\textbf{SNG}) and the \code{simple-err-fix} prompt (\textbf{SEF}). \dn{} indicates a percentage decrease relative to \textbf{FULL}, \up{} indicates a percentage increase, and \lr{} indicates no change. } 
    \label{tab:eff-harn-gen}
    \Description{}
\end{table*}

Within 10 generation attempts, \codename produced syntactically correct harnesses for all 9 Rust libraries. For \code{image-webp} and \code{prost-types}, every harness was correct on the first synthesis pass. With up to three generation attempts (\code{@Pass$\leq$3}), the overall syntactic correctness rate reached 97.3\%, with 5 out of 9 libraries achieving 100\%. In total, \codename generated \numofharns{} harnesses across the 9 libraries (32 per library on average), requiring an average of 46 generation attempts, or about 1.4 attempts per valid harness. These results highlight the effectiveness of harness generation.

\noindent{\textbf{Performance.}} For the \numofharns{} harnesses, \codename with GPT-4.1 achieved an average generation time of \avgtimegensec{} (\avgtimegen{}). This time included both compiler execution and network latency. The average cost per harness generation was \$0.03. Additional LLMs were evaluated in \autoref{sec:alternative_llms}.

\noindent{\textbf{Bug Discovery.}}
We ran Kani on the harnesses generated by \codename and identified six memory safety issues in \code{pdf-rs}, as shown in \autoref{tab:bugs}. Five issues have been fixed, while one is still under developer review. 

\begin{table}[htb]
    \arrayrulecolor{gray!100}
    {\footnotesize
    \begin{tabular}{l|l|c|c}
        \toprule
        \textbf{File} & \textbf{Buggy Function} & \textbf{Bug Type} & \textbf{Fixed} \\
        \midrule
        parse\_xref.rs & read\_u64\_from\_stream & Arith. Overflow & \checkmark \\
        font.rs & utf16be\_to\_string\_lossy & Access out-of-bound & \checkmark \\
        file.rs & load\_storage\_and\_trailer\_pwd & Access out-of-bound & \checkmark \\ 
        crypt.rs & Decoder::decrypt & Access out-of-bound  & \checkmark \\ 
        crypt.rs & Decoder::key & Access out-of-bound & $\times$ \\
        primitive.rs & Date::from\_primitive & Not char boundary  & \checkmark \\
        \bottomrule
    \end{tabular}
    }
    \caption{Bugs found in pdf-rs (version: 0.9.0).}
    \label{tab:bugs}
\end{table}

\autoref{fig:bug-func} demonstrates a bug occurred in function \code{utf16be\_to\_ string\_lossy}. It calls \code{utf16be\_to\_char}, which in turn invokes \code{char ::decode\_utf16} over an iterator that maps each 2-byte chunk of data to a \code{u16} using \code{u16::from\_be\_bytes([w[0],w[1]])}. When the length of \code{data} is less than 2, accessing \code{w[1]} results in an out-of-bound access.
\autoref{fig:bug-demo} shows a test function for the buggy function, the extracted \sketch{}, and the generated harness. The harness initialized a nondet vector whose length was also nondet. The LLM specified a maximum length, allowing the vector to have any size less than \code{SLICE\_MAX\_LEN}. This vector was then passed to the \sketch{} \code{scen\_test\_to\_char\_4}.
The bug was detected with ``\code{cargo kani {-}{-}tests {-}{-}harness harness\_scen\_test\_to\_char {-}{-}\\unwind 10 {-}{-}no-unwinding-checks}". Setting an unwind limit is generally essential for analyzing real-world codebases with intricate functions and loops, as it prevents BMC from exhausting resources. 

\begin{figure}[htb]
    \centering
    \begin{subfigure}{0.98\linewidth}
        \begin{ScriptSizePygex}[\tiny]
            \input{pygtex/bug-fn-1.pygtex}
        \end{ScriptSizePygex}
    \end{subfigure}
    \hfill
     \begin{subfigure}{0.98\linewidth}
        \begin{ScriptSizePygex}[\tiny]
            \input{pygtex/bug-fn-2.pygtex}
        \end{ScriptSizePygex}
    \end{subfigure}
    \caption{Buggy function with the dataflow highlighted.}
    \label{fig:bug-func}
    \Description{}
    % \vspace{-5ex}
\end{figure}

\begin{figure}[htb]
    \centering
    \begin{subfigure}{0.45\linewidth}
        \begin{ScriptSizePygex}[\tiny]
            \input{pygtex/test-to-char.pygtex}
        \end{ScriptSizePygex}
    \end{subfigure}
    \hfill
     \begin{subfigure}{0.54\linewidth}
        \begin{ScriptSizePygex}[\tiny]
            \input{pygtex/harn-fn.pygtex}
        \end{ScriptSizePygex}
    \end{subfigure}
    \caption{The harness for the buggy function, along with its test function and the extracted scenario function.}
    \label{fig:bug-demo}
    \Description{}
\end{figure}

% newly add sparse/weak tests
\noindent{\textbf{Sparse Tests.}} Harness generation success is unaffected by sparse tests. In 9 libraries, we randomly selected a limited number of existing test functions to simulate such cases (\code{\#Tests} shows the counts). As shown in ~\autoref{tab:sparse_weak_tests}, success rates remain close to 100\% within three attempts.
However, scenario coverage decreased under sparse tests. To address this, for projects with sparse tests, additional tests could be generated, for instance, by using LLMs to summarize function semantics and derive meaningful calling sequences to enrich scenarios before applying \codename.

\begin{table}[htb]
    \centering
    \arrayrulecolor{gray!100}
    {\footnotesize
    \begin{tabular}{c|c|c|c|c|c}
    \toprule
         \textbf{Libray} & \textbf{@Pass=0} & \textbf{@Pass$\leq$3} & \textbf{@Pass$\leq$5} & \textbf{@Pass$\leq$10} & \textbf{\#Tests} \\
         \midrule
         pdf-rs & 71\% & 100\% & 100\% & 100\% & 6 \\
         tar-rs & 60\% & 100\% & 100\% & 100\% & 10 \\
         jpeg-decoder & 50\% & 100\% & 100\% & 100\% & 2 \\ 
         tempfile & 75\% & 75\% & 100\%	& 100\% & 5 \\
         jzon-rs & 70\% & 100\% & 100\% & 100\% & 8 \\
         image-webp & 100\% & 100\% & 100\% & 100\% & 3 \\
         lexical-util & 80\% & 100\% & 100\% & 100\% & 3 \\
         prost-types & 100\% & 100\% & 100\% & 100\% & 3 \\
         p256 & 50\% & 100\% & 100\% & 100\% & 4 \\
         \hline 
         \textbf{Average} & 72.9\% & 97.2\% & 100\% & 100\% & 5 \\
         \bottomrule
    \end{tabular}
    }
    \caption{Harness generation success rate on sparse tests with the complete workflow (FULL).}
    \label{tab:sparse_weak_tests}
\end{table}

% \begin{table*}[htb]
%     \centering
%     \arrayrulecolor{gray!100}
%     {\footnotesize
%     \begin{tabular}{c|c|c|c|c|c|c}
%     \toprule
%          \textbf{Libray} & \textbf{\#Harn.} & \textbf{@Pass=0} & \textbf{@Pass$\leq$3} & \textbf{@Pass$\leq$5} & \textbf{@Pass$\leq$10} & \textbf{\#Generations} \\
%          \midrule
%          pdf-rs & 27 & 63\% & 93\% & 97\% & 100\% & 46 \\
%          tar-rs & 82 & 70\% & 98\% & 100\% & 100\% & 137 \\
%          jpeg-decoder & 7 & 86\% & 100\% & 100\% & 100\% & 11 \\ 
%          tempfile & 36 & 81\% & 100\% & 100\%	& 100\% & 44 \\
%          jzon-rs & 80 & 83\% & 95\% & 96\% & 100\% & 107 \\
%          image-webp & 9 & 100\% & 100\% & 100\% & 100\% & 9 \\
%          lexical-util & 27 & 63\% & 100\% & 100\% & 100\% & 37 \\
%          prost-types & 6 & 100\% & 100\% & 100\% & 100\% & 6 \\
%          p256 & 75\% & 20 & 90\% & 100\% & 100\% & 31 \\
%          \hline 
%          \textbf{Average} & 32 & 80.6\% & 97.3\% & 97.8\% & 100\% & 46 \\
%          \bottomrule
%     \end{tabular}
%     }
%     \caption{Harness generation success rate with the complete workflow (FULL).}
%     \label{tab:xxx}
% \end{table*}

\subsection{RQ2: Ablation Study}~\label{sec:ablation-study}
For harness synthesis, we use a dependency graph and incorporate external knowledge to generate CoT instructions, enabling LLMs to incrementally construct nondet parameters. During harness compilation, we guide LLMs' attention on generating reasonable fixes. This section evaluates the contributions of these methods to the successful generation of harnesses with the dataset $D_{scen}$.

\noindent\textbf{Contribution of CoT Instructions Design.}
We designed a new prompt, \code{simple-nondet-gen}, for a comparative experiment. This prompt differs from the original shown in \autoref{fig:gen_harn_for_sketch_fn} by removing the highlighted ``Nondet Argument Construction" section containing the CoT instructions.
Column \code{SNG} in \autoref{tab:eff-harn-gen} presents the results using this new prompt. 
Column \code{@Pass=0} shows the proportion of harnesses that were syntactically correct immediately after the harness synthesis stage. 
Compared to the original prompt (Column \code{FULL}), the absence of nondet generation guidance resulted in an average decrease of 23.1\% in the proportion of correctly synthesized harnesses. Specifically, \code{pdf-rs} saw the largest decline (37\%), as its \sketchs{} involved more user-defined parameter types. In the subsequent iterative error-fixing phases, the proportions of successfully generated harnesses decreased by 3.3\%, 1.9\%, and 1.0\% at \code{@Pass$\leq$3,5,10}, respectively.

Without the CoT guidance, the LLM needed more attempts to produce correct harnesses. For example, the number of generation attempts for \code{lexical-util} increased by over 54\%. When the interaction limit was set to 10, the total number of generation attempts by the LLM increased by an average of 28.3\%.

\noindent\textbf{Contribution of Error-Fix Design.}
We used \code{simple-err-fix} (\code{SEF)} shown in \autoref{fig:err_feedback_simple} for comparative evaluations, and \autoref{tab:eff-harn-gen} presents the results. With \code{SEF}, the first-pass success rate for producing syntactically correct harnesses dropped by 18.7\% compared to the full workflow. \code{prost-types} suffered the largest decline (67\%), while \code{jpeg-decoder}, \code{tempfile}, and \code{image-webp} showed no decrease. In the subsequent multi-round error-fixing phases, the success rates at \code{Pass$\leq$3,5,10} fell by 7.7\%, 0.8\%, and 0.4\%, respectively. Overall, the average number of generation attempts increased by 23.9\%.

\subsection{RQ3: Comparisons}
Autoharness~\cite{autoharness}, developed by Kani~\cite{kani}, generates nondeterministic parameters by automatically creating inputs for types that implement \code{Kani::Arbitrary} trait and invoking the target functions with these inputs. For Rust primitive types like integers, it automatically generates nondet data, thereby eliminating the need for tedious manual workload. 

\begin{figure}[htb]
    \centering
    \includegraphics[width=0.95\linewidth]{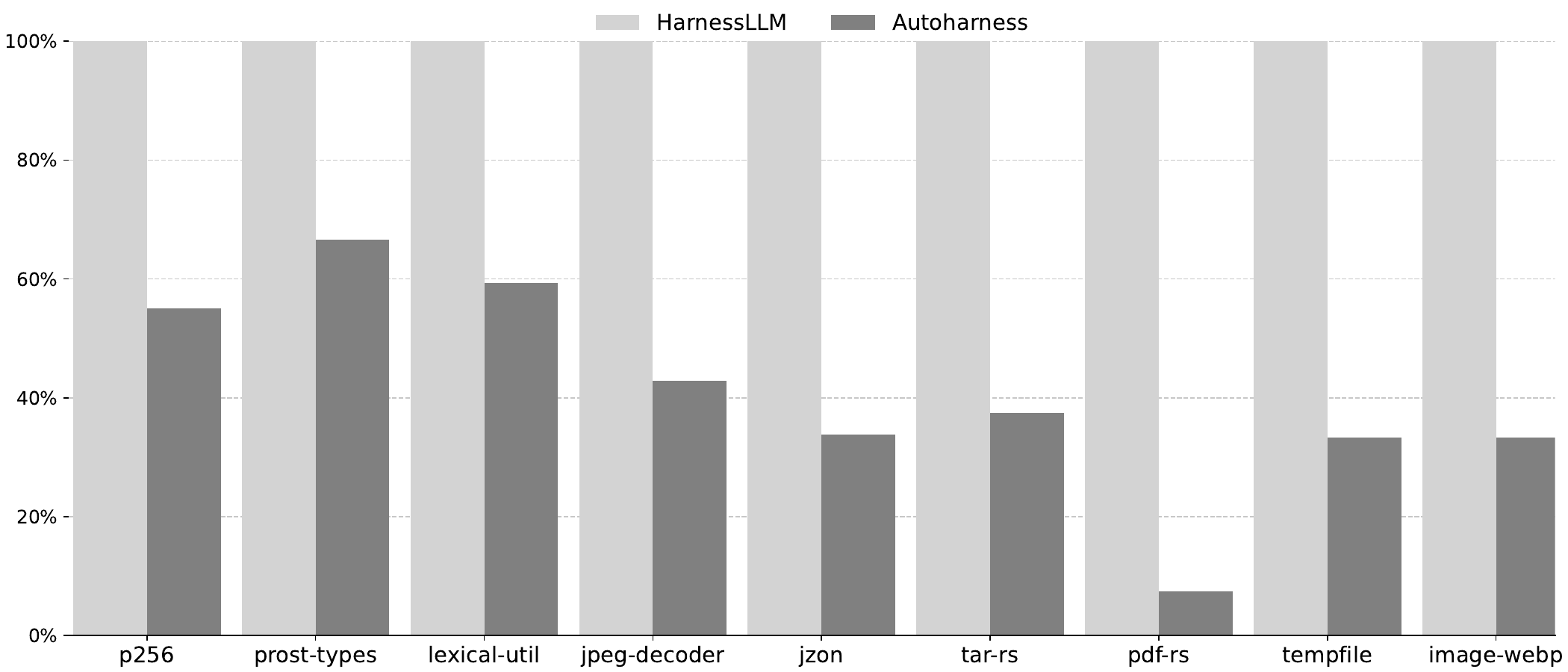}
    \caption{Comparison with Autoharness on $D_{scen}$. The average generation success rate of Autoharness was \autoharnsuprate{}, while \codename was 100\%.}
    \label{fig:comp_autoharness}
    \Description{}
\end{figure}

To ensure a fair comparison, we provided the \sketchs{} from $D_{scen}$ to both Autoharness and \codename, and compared the number of successfully generated harnesses. \autoref{fig:comp_autoharness} illustrates their harness generation performance across various libraries. For \code{pdf-rs}, Autoharness failed to generate harnesses for 92.59\% of the \sketchs{}. The best performance was observed in \code{prost-types}, where Autoharness generated harnesses for around 66.67\% of the \sketchs{}. The average generation success rate was \autoharnsuprate{}. 
This limitation arises because Autoharness currently supports nondet generation only for basic types that implement \code{kani::Arbitrary} trait and cannot handle complex types, particularly user-defined ones within the crate.
In contrast, \codename has no such restrictions. By generating CoT instructions from the dependency graph, \codename successfully constructs the necessary nondet parameters, enabling harness generation for every \sketch{}.

\subsection{RQ4: Alternative Models}
\label{sec:alternative_llms}
\begin{figure}[htb]
    \centering
    \includegraphics[width=0.95\linewidth]{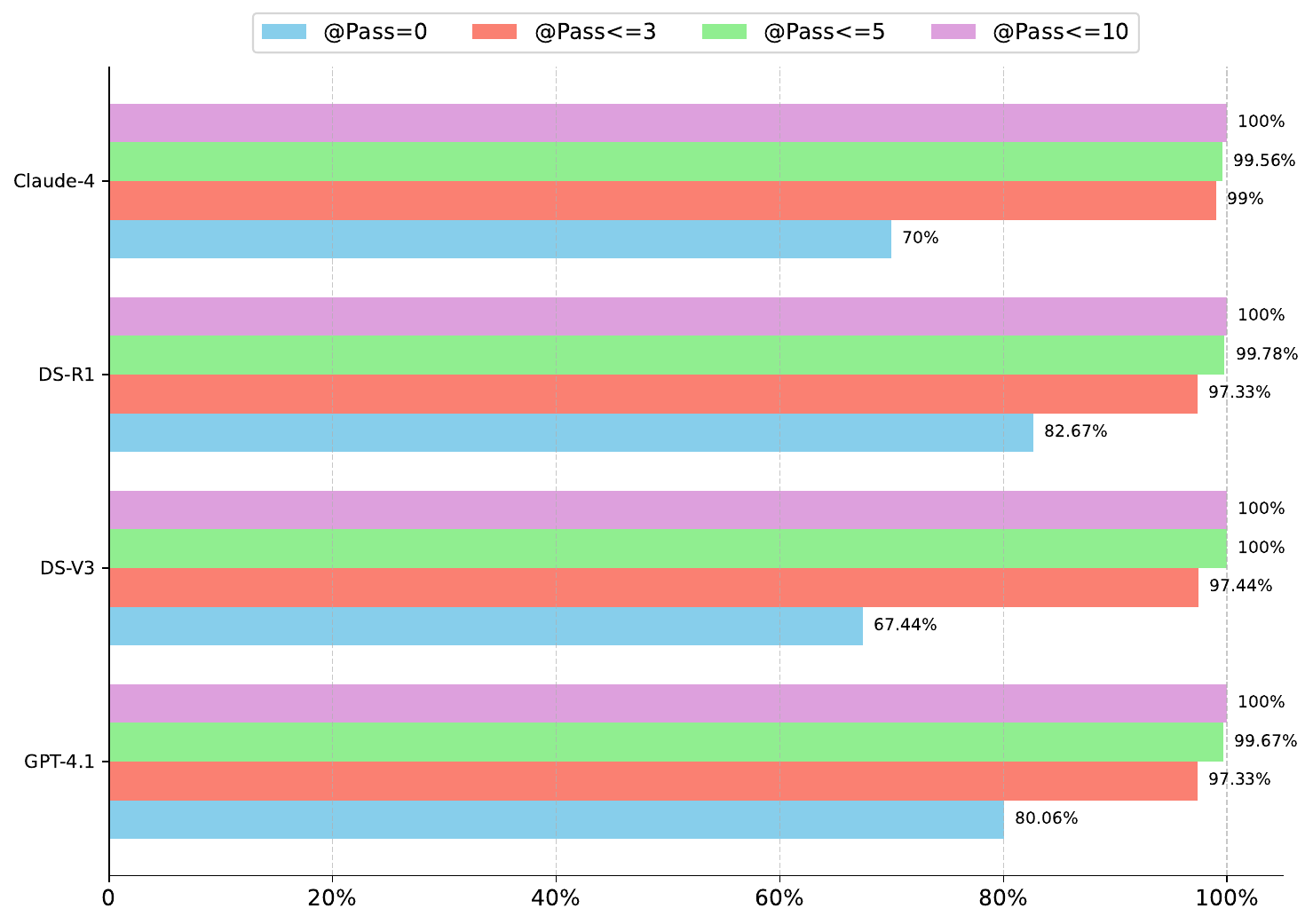}
    \caption{Comparison of LLMs on generation success rate within 10 attempts.}
    \label{fig:comp_llms_harn_gen}
    \Description{}
\end{figure}

\autoref{fig:comp_llms_harn_gen} presents the generation results of \codename on the $D_{rwd}$ dataset using additional LLMs.
All models achieved a 100\% success rate within 10 attempts. Specifically, in the first generation round, the reasoning model DS-R1 (Deepseek-R1-0528) led with a success rate of 82.67\%, followed by Claude-4 (claude-sonnet-4-20250514) at 70\%, and DS-V3 (DeepSeek-V3-0324) with the lowest at 67.44\%. The initial harness generation relies on the LLM's understanding of the guidelines for nondet generation, highlighting DS-R1's superior comprehension capabilities in this context.

\begin{table}[htb]
    \centering
    {\small
    \begin{tabular}{c|c|c|c|c}
        \midrule
        LLMs & GPT-4.1 & DS-V3 & Claude-4 & DS-R1 \\
        \hline
        Avg. Time (sec) & 145 & 353 & 188 & 1526 \\
        \hline
        Avg. Cost (USD) & 0.03 & 0.004 & 0.05 & 0.04 \\
        \bottomrule
    \end{tabular}
    }
    \caption{Comparison of generation time and cost per harness.}
    \label{tab:comp_time_costs}
\end{table}

\autoref{tab:comp_time_costs} presents the differences among various LLMs in terms of generation time and token cost per harness. Claude-4 incurred the highest average token cost (\$0.05), while DS-V3 had the lowest (\$0.004). DS-R1 exhibited the longest average harness generation time, as it was the only reasoning model tested, dedicating a significant portion of its time to generating its reasoning steps. In contrast, GPT-4.1 achieved the shortest average generation time.

%% file: discussion.tex
\section{Discussion}
\noindent\textbf{Traditional Approaches vs. LLMs.}
Traditional program analysis approaches like custom LLVM~\cite{smack} or MIR passes~\cite{mirai, mir-checker}, can address parts of our pipeline but often struggle with Rust’s advanced features (e.g., generics, closures, and higher‐order functions) and are prone to breaking across different Rustc or LLVM versions. By contrast, LLMs excel at code analysis and generation, handling these complexities through well-designed natural language prompts. Therefore, we propose integrating LLMs into our workflow to provide a lightweight but effective solution.

\noindent\textbf{Difference from LLM-based Fuzzing Harness Generation.}
Unlike prior LLM-based harness generation for fuzzers (e.g., ~\cite{promptfuzz}, ~\cite{ckgfuzz}), which often relies on type dependencies or unconstrained LLM predictions and thus suffers from API misuse, our approach extracts realistic calling scenarios directly from existing well-crafted tests by developers, greatly mitigating misuse issues. Moreover, \codename is tailored for Rust. It accounts for language-specific features such as traits and leverages Kani knowledge, whereas existing fuzzing harness methods have limited applicability in this setting.

\noindent\textbf{Coverage.}
The coverage of calling scenarios in the generated harnesses depends on the comprehensiveness of the existing test cases. Our method achieves high success in preserving scenarios, as test cases are typically well-designed by developers. If certain functions lack coverage, additional test cases can be generated to enhance scenario coverage. Furthermore, the Kani harnesses we create use unconstrained symbolic variables for arguments, so the statement coverage can be guaranteed.

\noindent\textbf{Bound Values.}
Kani, as a BMC, requires explicit bounds for slice lengths. LLMs infer suitable bounds, either constant or unconstrained, based on code understanding, as demonstrated in~\autoref{fig:bug-demo}. Providing appropriate bounds remains a challenge for BMCs. Interval analysis~\cite{ermedahl2007loop, unsafecop} could help determine bounds. Moreover, as Kani reports unwinding errors along with runtime execution contexts when programs fail to fully unwind, supplying this information, together with relevant code contexts, to LLMs could enable adaptive bound selections. We leave this for future work. 

\noindent\textbf{Function Tracing.}
\codename only instruments functions in surface Rust code, potentially missing those generated by macros not visible at this level. However, macros make up a small portion of most Rust codebases, so this limitation has minimal impact on the results. As all the functions are visible in Rust's MIR or LLVM IR, we plan to extend instrumentation to those levels by writing passes in future work.

%% file: related.tex
\section{Related Work}
\noindent\textbf{Rust Verification.}
Several studies focus on Rust verification.
Theorem proving approaches like RustBelt~\cite{rustbelt}, Aeneas~\cite{aeneas}, Refined Rust~\cite{refined-rust}, and HAX~\cite{hax} translate Rust’s MIR into Coq or F* to establish type-system soundness. Prusti~\cite{prusti} and Creusot~\cite{creusot} apply deductive verification to safe Rust by requiring user‐written function contracts and loop invariants. Verus~\cite{verus} uses SMT‐based proofs to verify safe Rust and certain unsafe constructs like raw pointers and RefCell. Gillian‐Rust~\cite{gillian-rust} combines automated verification for safe Rust with separation‐logic to handle unsafe code, eliminating the need for external harnesses.

Automatic verification techniques like BMC~\cite{kani, smack} and symbolic execution~\cite{rvt, crux, panic-checker-klee} rely on explicit harnesses. 
Smack~\cite{smack} translates LLVM bitcode to Boogie IR~\cite{boogie} to detect memory safety bugs in unsafe Rust. Kani~\cite{kani} checks both safe and unsafe code, verifying a subset of Rust’s undefined behaviors and user assertions, but cannot guarantee unbounded proof. UnsafeCop~\cite{unsafecop} extends Kani with loop bound inference, loop stubbing and scheduling strategies to improve scalability. 
To reduce the manual burden of harness writing, Autoharness~\cite{autoharness} automates harness generation for functions whose parameters implement \code{kani::Arbitrary}, but it cannot handle user-defined types. PropProof~\cite{propproof} converts existing proptest~\cite{proptest} harnesses into Kani harnesses. TraitInv~\cite{traitinv} synthesizes harnesses for some built-in traits but not user-defined ones. Erdin~\cite{erdin} targets user-defined correctness properties but requires developer-supplied annotations to generate harnesses.

\noindent\textbf{LLM for Harness Generation.}
No prior work has directly applied LLMs to generate verification harnesses, but related studies have employed LLMs to generate testing harnesses, such as fuzzing harnesses~\cite{ckgfuzz, promptfuzz, whitefox}, and standard test cases~\cite{symprompt,codemosa,jnit-test-empirical-study}. GPTFuzz~\cite{gptfuzz} leverages LLMs to generate vulnerable inputs for assessing the robustness of deep learning library APIs. PromptFuzz~\cite{promptfuzz} introduces an iterative fuzzing loop that generates drivers to explore previously untested code paths. CKGFuzzer~\cite{ckgfuzz} uses a code knowledge graph via interprocedural analysis to create fuzz drivers. Whitefox~\cite{whitefox} generates test programs targeting deep learning compilers to uncover optimization bugs. 
Several works also explore unit test generation with LLMs. CodaMosa~\cite{codemosa} supplies test cases for uncovered functions when search-based methods reach coverage saturation. ChatUnitTest~\cite{xie-chatunitest} generates unit tests by extracting key project information and building an adaptive focal context that fits within the LLM’s token limit. Tang et al.~\cite{tang-chatgpt-vs-sbst} systematically compared test suites generated by ChatGPT with those from state-of-the-art search-based software testing tools.

\noindent\textbf{LLM for Rust Verification.}
Another related research~\cite{jiannan-yao-proof-syntheis, autoverus, safe} explore the integration of LLMs with Rust verification, focusing on automatically generating the proofs necessary for program correctness. The work~\cite{jiannan-yao-proof-syntheis} decomposes the verification process into smaller tasks by iteratively querying LLMs and combining their outputs with lightweight static analysis, thereby synthesizing proof structures such as function contracts, invariants, and assertions for Verus and significantly reducing the human workload. AutoVerus~\cite{autoverus} integrates expert knowledge with formal methods to assist LLMs in generating proofs, employing LLM agents to perform preliminary proof generation, refine proofs based on general guidelines, and debug proofs through verification errors, achieving a 90\% success rate in producing corrected proofs. Similarly, SAFE~\cite{safe} introduces a self-evolving framework that addresses data scarcity by combining data synthesis with model fine-tuning, demonstrating superior efficiency and precision over approaches that rely solely on GPT-4.

%% file: conclusion.tex
\section{Conclusion}
We introduce \codename, an automated workflow leveraging LLMs to generate verification harnesses for Rust code directly from existing test suites. It extracts calling scenarios from test cases, constructs harnesses with nondeterministic arguments, and iteratively refines them using compiler feedback. In evaluations across \numofcrates{} real-world Rust codebases, \codename extracted \numofsketchs{} calling scenarios from \teststotalnumber{} test cases with a precision of \scenpresvrate{}, successfully generated \numofharns{} harnesses, achieving a 100\% success rate, with an average generation time of around \avgtimegensec{} per harness, outperforming Autoharness which succeeded on only \autoharnsuprate{} of the extracted scenarios. Finally, 6 real-world memory safety bugs were identified using the generated harnesses, demonstrating the practical utility of our approach. To the best of our knowledge, \codename is the first tool to use LLMs for generating harnesses aimed at memory safety verification in real-world Rust projects.

%% file: sample-base.bib
@String{Computing = "Computing" }

@String{Computer = "{IEEE} Computer" }

@String{Springer = "Springer-Verlag" }

@InProceedings{unsafecop,
author="Wang, Minghua
and Xue, Jingling
and Huang, Lin
and Zi, Yuan
and Wei, Tao",
editor="Platzer, Andre
and Rozier, Kristin Yvonne
and Pradella, Matteo
and Rossi, Matteo",
title="UnsafeCop: Towards Memory Safety for Real-World Unsafe Rust Code with Practical Bounded Model Checking",
booktitle="Formal Methods",
year="2025",
publisher="Springer Nature Switzerland",
address="Cham",
pages="307--324",
isbn="978-3-031-71177-0"
}

@article{aeneas,
author = {Ho, Son and Protzenko, Jonathan},
title = {Aeneas: Rust verification by functional translation},
year = {2022},
issue_date = {August 2022},
publisher = {Association for Computing Machinery},
address = {New York, NY, USA},
volume = {6},
number = {ICFP},
url = {https://doi.org/10.1145/3547647},
doi = {10.1145/3547647},
journal = {Proc. ACM Program. Lang.},
month = aug,
articleno = {116},
numpages = {31}
}

@misc{hax,
      author = {Karthikeyan Bhargavan and Maxime Buyse and Lucas Franceschino and Lasse Letager Hansen and Franziskus Kiefer and Jonas Schneider-Bensch and Bas Spitters},
      title = {hax: Verifying Security-Critical Rust Software using Multiple Provers},
      howpublished = {Cryptology {ePrint} Archive, Paper 2025/142},
      year = {2025},
      url = {https://eprint.iacr.org/2025/142}
}

@inproceedings{kani,
author = {VanHattum, Alexa and Schwartz-Narbonne, Daniel and Chong, Nathan and Sampson, Adrian},
title = {Verifying dynamic trait objects in rust},
year = {2022},
isbn = {9781450392266},
publisher = {Association for Computing Machinery},
address = {New York, NY, USA},
url = {https://doi.org/10.1145/3510457.3513031},
doi = {10.1145/3510457.3513031},
booktitle = {Proceedings of the 44th International Conference on Software Engineering: Software Engineering in Practice},
pages = {321–330},
numpages = {10},
location = {Pittsburgh, Pennsylvania},
series = {ICSE-SEIP '22}
}

@inproceedings{verus,
author = {Lattuada, Andrea and Hance, Travis and Bosamiya, Jay and Brun, Matthias and Cho, Chanhee and LeBlanc, Hayley and Srinivasan, Pranav and Achermann, Reto and Chajed, Tej and Hawblitzel, Chris and Howell, Jon and Lorch, Jacob R. and Padon, Oded and Parno, Bryan},
title = {Verus: A Practical Foundation for Systems Verification},
year = {2024},
isbn = {9798400712517},
publisher = {Association for Computing Machinery},
address = {New York, NY, USA},
url = {https://doi.org/10.1145/3694715.3695952},
doi = {10.1145/3694715.3695952},
booktitle = {Proceedings of the ACM SIGOPS 30th Symposium on Operating Systems Principles},
pages = {438–454},
numpages = {17},
location = {Austin, TX, USA},
series = {SOSP '24}
}

@InProceedings{prusti,
author="Astrauskas, Vytautas
and B{\'i}l{\'y}, Aurel
and Fiala, Jon{\'a}{\v{s}}
and Grannan, Zachary
and Matheja, Christoph
and M{\"u}ller, Peter
and Poli, Federico
and Summers, Alexander J.",
editor="Deshmukh, Jyotirmoy V.
and Havelund, Klaus
and Perez, Ivan",
title="The Prusti Project: Formal Verification for Rust",
booktitle="NASA Formal Methods",
year="2022",
publisher="Springer International Publishing",
address="Cham",
pages="88--108",
isbn="978-3-031-06773-0"
}

@InProceedings{creusot,
author="Denis, Xavier
and Jourdan, Jacques-Henri
and March{\'e}, Claude",
editor="Riesco, Adrian
and Zhang, Min",
title="Creusot: A Foundry for the Deductive Verification of Rust Programs",
booktitle="Formal Methods  and Software Engineering",
year="2022",
publisher="Springer International Publishing",
address="Cham",
pages="90--105",
isbn="978-3-031-17244-1"
}

@InProceedings{smack,
author="Rakamari{\'{c}}, Zvonimir
and Emmi, Michael",
editor="Biere, Armin
and Bloem, Roderick",
title="SMACK: Decoupling Source Language Details from Verifier Implementations",
booktitle="Computer Aided Verification",
year="2014",
publisher="Springer International Publishing",
address="Cham",
pages="106--113",
isbn="978-3-319-08867-9"
}

@misc{crux,
      title={Crux, a Precise Verifier for Rust and Other Languages}, 
      author={Stuart Pernsteiner and Iavor S. Diatchki and Robert Dockins and Mike Dodds and Joe Hendrix and Tristan Ravich and Patrick Redmond and Ryan Scott and Aaron Tomb},
      year={2024},
      eprint={2410.18280},
      archivePrefix={arXiv},
      primaryClass={cs.PL},
      url={https://arxiv.org/abs/2410.18280}, 
}

@article{refined-rust,
author = {G\"{a}her, Lennard and Sammler, Michael and Jung, Ralf and Krebbers, Robbert and Dreyer, Derek},
title = {RefinedRust: A Type System for High-Assurance Verification of Rust Programs},
year = {2024},
issue_date = {June 2024},
publisher = {Association for Computing Machinery},
address = {New York, NY, USA},
volume = {8},
number = {PLDI},
url = {https://doi.org/10.1145/3656422},
doi = {10.1145/3656422},
journal = {Proc. ACM Program. Lang.},
month = jun,
articleno = {192},
numpages = {25}
}

@inproceedings{panic-checker-klee,
author = {Zhang, Ying and Li, Peng and Ding, Yu and Wang, Lingxiang and Williams, Dan and Meng, Na},
title = {Broadly Enabling KLEE to Effortlessly Find Unrecoverable Errors in Rust},
year = {2024},
isbn = {9798400705014},
publisher = {Association for Computing Machinery},
address = {New York, NY, USA},
url = {https://doi.org/10.1145/3639477.3639714},
doi = {10.1145/3639477.3639714},
booktitle = {Proceedings of the 46th International Conference on Software Engineering: Software Engineering in Practice},
pages = {441–451},
numpages = {11},
location = {Lisbon, Portugal},
series = {ICSE-SEIP '24}
}

@misc{gillian-rust,
      title={A hybrid approach to semi-automated Rust verification}, 
      author={Sacha-Élie Ayoun and Xavier Denis and Petar Maksimović and Philippa Gardner},
      year={2025},
      eprint={2403.15122},
      archivePrefix={arXiv},
      primaryClass={cs.PL},
      url={https://arxiv.org/abs/2403.15122}, 
}

@article{rustbelt,
  title={RustBelt: Securing the foundations of the Rust programming language},
  author={Jung, Ralf and Jourdan, Jacques-Henri and Krebbers, Robbert and Dreyer, Derek},
  journal={Proceedings of the ACM on Programming Languages},
  volume={2},
  number={POPL},
  pages={1--34},
  year={2017},
  publisher={ACM New York, NY, USA}
}

@article{boogie,
  title={This is boogie 2},
  author={Leino, K Rustan M},
  journal={manuscript KRML},
  volume={178},
  number={131},
  pages={9},
  year={2008},
  publisher={Citeseer}
}

@misc{proptest,
  title = "Proptest",
  author = "The proptest developers",
  url = "https://github.com/proptest-rs/proptest",
  year = "2025", 
  note = "Last accessed Nov.2025"
}

@inproceedings{propproof,
author = {Takashima, Yoshiki},
title = {PropProof: Free Model-Checking Harnesses from PBT},
year = {2023},
isbn = {9798400703270},
publisher = {Association for Computing Machinery},
address = {New York, NY, USA},
url = {https://doi.org/10.1145/3611643.3613863},
doi = {10.1145/3611643.3613863},
booktitle = {Proceedings of the 31st ACM Joint European Software Engineering Conference and Symposium on the Foundations of Software Engineering},
pages = {1903–1913},
numpages = {11},
location = {San Francisco, CA, USA},
series = {ESEC/FSE 2023}
}

@InProceedings{traitinv,
author="Byrnes, Twain
and Takashima, Yoshiki
and Jia, Limin",
editor="Dimitrova, Rayna
and Lahav, Ori
and Wolff, Sebastian",
title="Automatically Enforcing Rust Trait Properties",
booktitle="Verification, Model Checking, and Abstract Interpretation",
year="2024",
publisher="Springer Nature Switzerland",
address="Cham",
pages="210--223",
isbn="978-3-031-50521-8"
}

@MASTERSTHESIS{erdin,
 author = {Matthias Erdin},
 title = {Verification of Rust Generics, Typestates, and Traits (Master's thesis)},
 year = {2019},
 source = {https://ethz.ch/content/dam/ethz/special-interest/infk/chair-program-method/pm/documents/Education/Theses/Matthias_Erdin_MA_report.pdf},
 school = {ETH Z¨urich},
 }

@MASTERSTHESIS{electrolysis,
 author = {Sebastian Ullrich},
 title = {Simple Verification of Rust Programs via Functional Purification (Master's thesis)},
 year = {2016},
 source = {https://pp.ipd.kit.edu/uploads/publikationen/ullrich16masterarbeit.pdf},
 school = {Fakultät für Informatik}
}

@misc{tree-sitter,
    author = "The Tree-sitter developer",
    url = "https://github.com/tree-sitter/tree-sitter",
    year = "2025",
    note = "Last accessed Nov. 2025"
}

@misc{logfn,
    author = "The logfn developer",
    url = "https://crates.io/crates/logfn",
    year = "2025",
    note = "Last accessed Nov. 2025"
}

@misc{mirai,
    author = "The Mirai developers",
    title = "MIRAI: Rust mid-level IR Abstract Interpreter",
    url = "https://github.com/facebookexperimental/MIRA",
    year = "2025",
    note = "Last accessed Nov. 2025",
}

@article{safedrop,
author = {Cui, Mohan and Chen, Chengjun and Xu, Hui and Zhou, Yangfan},
title = {SafeDrop: Detecting Memory Deallocation Bugs of Rust Programs via Static Data-flow Analysis},
year = {2023},
issue_date = {July 2023},
publisher = {Association for Computing Machinery},
address = {New York, NY, USA},
volume = {32},
number = {4},
issn = {1049-331X},
url = {https://doi.org/10.1145/3542948},
doi = {10.1145/3542948},
journal = {ACM Trans. Softw. Eng. Methodol.},
month = may,
articleno = {82},
numpages = {21}
}

@inproceedings{mir-checker,
author = {Li, Zhuohua and Wang, Jincheng and Sun, Mingshen and Lui, John C.S.},
title = {MirChecker: Detecting Bugs in Rust Programs via Static Analysis},
year = {2021},
isbn = {9781450384544},
publisher = {Association for Computing Machinery},
address = {New York, NY, USA},
url = {https://doi.org/10.1145/3460120.3484541},
doi = {10.1145/3460120.3484541},
booktitle = {Proceedings of the 2021 ACM SIGSAC Conference on Computer and Communications Security},
pages = {2183–2196},
numpages = {14},
location = {Virtual Event, Republic of Korea},
series = {CCS '21}
}

@inproceedings{rudra,
author = {Bae, Yechan and Kim, Youngsuk and Askar, Ammar and Lim, Jungwon and Kim, Taesoo},
title = {Rudra: Finding Memory Safety Bugs in Rust at the Ecosystem Scale},
year = {2021},
isbn = {9781450387095},
publisher = {Association for Computing Machinery},
address = {New York, NY, USA},
url = {https://doi.org/10.1145/3477132.3483570},
doi = {10.1145/3477132.3483570},
booktitle = {Proceedings of the ACM SIGOPS 28th Symposium on Operating Systems Principles},
pages = {84–99},
numpages = {16},
location = {Virtual Event, Germany},
series = {SOSP '21}
}

@article{yuga,
author = {Nitin, Vikram and Mulhern, Anne and Arora, Sanjay and Ray, Baishakhi},
title = {Yuga: Automatically Detecting Lifetime Annotation Bugs in the Rust Language},
year = {2024},
issue_date = {Oct. 2024},
publisher = {IEEE Press},
volume = {50},
number = {10},
issn = {0098-5589},
url = {https://doi.org/10.1109/TSE.2024.3447671},
doi = {10.1109/TSE.2024.3447671},
journal = {IEEE Trans. Softw. Eng.},
month = oct,
pages = {2602–2613},
numpages = {12}
}

@article{tang-chatgpt-vs-sbst,
author = {Tang, Yutian and Liu, Zhijie and Zhou, Zhichao and Luo, Xiapu},
title = {ChatGPT vs SBST: A Comparative Assessment of Unit Test Suite Generation},
year = {2024},
issue_date = {June 2024},
publisher = {IEEE Press},
volume = {50},
number = {6},
issn = {0098-5589},
url = {https://doi.org/10.1109/TSE.2024.3382365},
doi = {10.1109/TSE.2024.3382365},
journal = {IEEE Trans. Softw. Eng.},
month = jun,
pages = {1340–1359},
numpages = {20}
}

@misc{gptfuzz,
      title={GPTFUZZER: Red Teaming Large Language Models with Auto-Generated Jailbreak Prompts}, 
      author={Jiahao Yu and Xingwei Lin and Zheng Yu and Xinyu Xing},
      year={2024},
      eprint={2309.10253},
      archivePrefix={arXiv},
      primaryClass={cs.AI},
      url={https://arxiv.org/abs/2309.10253}, 
}

@inproceedings{promptfuzz,
author = {Lyu, Yunlong and Xie, Yuxuan and Chen, Peng and Chen, Hao},
title = {Prompt Fuzzing for Fuzz Driver Generation},
year = {2024},
isbn = {9798400706363},
publisher = {Association for Computing Machinery},
address = {New York, NY, USA},
url = {https://doi.org/10.1145/3658644.3670396},
doi = {10.1145/3658644.3670396},
booktitle = {Proceedings of the 2024 on ACM SIGSAC Conference on Computer and Communications Security},
pages = {3793–3807},
numpages = {15},
location = {Salt Lake City, UT, USA},
series = {CCS '24}
}

@misc{ckgfuzz,
      title={CKGFuzzer: LLM-Based Fuzz Driver Generation Enhanced By Code Knowledge Graph}, 
      author={Hanxiang Xu and Wei Ma and Ting Zhou and Yanjie Zhao and Kai Chen and Qiang Hu and Yang Liu and Haoyu Wang},
      year={2024},
      eprint={2411.11532},
      archivePrefix={arXiv},
      primaryClass={cs.SE},
      url={https://arxiv.org/abs/2411.11532}, 
}

@inproceedings{codemosa,
author = {Lemieux, Caroline and Inala, Jeevana Priya and Lahiri, Shuvendu K. and Sen, Siddhartha},
title = {CodaMosa: Escaping Coverage Plateaus in Test Generation with Pre-Trained Large Language Models},
year = {2023},
isbn = {9781665457019},
publisher = {IEEE Press},
url = {https://doi.org/10.1109/ICSE48619.2023.00085},
doi = {10.1109/ICSE48619.2023.00085},
booktitle = {Proceedings of the 45th International Conference on Software Engineering},
pages = {919–931},
numpages = {13},
address = {Melbourne, Victoria, Australia},
series = {ICSE '23}
}

@article{whitefox,
author = {Yang, Chenyuan and Deng, Yinlin and Lu, Runyu and Yao, Jiayi and Liu, Jiawei and Jabbarvand, Reyhaneh and Zhang, Lingming},
title = {WhiteFox: White-Box Compiler Fuzzing Empowered by Large Language Models},
year = {2024},
issue_date = {October 2024},
publisher = {Association for Computing Machinery},
address = {New York, NY, USA},
volume = {8},
number = {OOPSLA2},
url = {https://doi.org/10.1145/3689736},
doi = {10.1145/3689736},
journal = {Proc. ACM Program. Lang.},
month = oct,
articleno = {296},
numpages = {27}
}

@inproceedings{xie-chatunitest,
author = {Chen, Yinghao and Hu, Zehao and Zhi, Chen and Han, Junxiao and Deng, Shuiguang and Yin, Jianwei},
title = {ChatUniTest: A Framework for LLM-Based Test Generation},
year = {2024},
isbn = {9798400706585},
publisher = {Association for Computing Machinery},
address = {New York, NY, USA},
url = {https://doi.org/10.1145/3663529.3663801},
doi = {10.1145/3663529.3663801},
booktitle = {Companion Proceedings of the 32nd ACM International Conference on the Foundations of Software Engineering},
pages = {572–576},
numpages = {5},
location = {Porto de Galinhas, Brazil},
series = {FSE 2024}
}

@article{symprompt,
author = {Ryan, Gabriel and Jain, Siddhartha and Shang, Mingyue and Wang, Shiqi and Ma, Xiaofei and Ramanathan, Murali Krishna and Ray, Baishakhi},
title = {Code-Aware Prompting: A Study of Coverage-Guided Test Generation in Regression Setting using LLM},
year = {2024},
issue_date = {July 2024},
publisher = {Association for Computing Machinery},
address = {New York, NY, USA},
volume = {1},
number = {FSE},
url = {https://doi.org/10.1145/3643769},
doi = {10.1145/3643769},
journal = {Proc. ACM Softw. Eng.},
month = jul,
articleno = {43},
numpages = {21}
}

@inproceedings{jnit-test-empirical-study,
author = {Siddiq, Mohammed Latif and Da Silva Santos, Joanna Cecilia and Tanvir, Ridwanul Hasan and Ulfat, Noshin and Al Rifat, Fahmid and Carvalho Lopes, Vin\'{\i}cius},
title = {Using Large Language Models to Generate JUnit Tests: An Empirical Study},
year = {2024},
isbn = {9798400717017},
publisher = {Association for Computing Machinery},
address = {New York, NY, USA},
url = {https://doi.org/10.1145/3661167.3661216},
doi = {10.1145/3661167.3661216},
booktitle = {Proceedings of the 28th International Conference on Evaluation and Assessment in Software Engineering},
pages = {313–322},
numpages = {10},
location = {Salerno, Italy},
series = {EASE '24}
}

@inproceedings{llm-se-android,
author = {Feng, Sidong and Chen, Chunyang},
title = {Prompting Is All You Need: Automated Android Bug Replay with Large Language Models},
year = {2024},
isbn = {9798400702174},
publisher = {Association for Computing Machinery},
address = {New York, NY, USA},
url = {https://doi.org/10.1145/3597503.3608137},
doi = {10.1145/3597503.3608137},
booktitle = {Proceedings of the IEEE/ACM 46th International Conference on Software Engineering},
articleno = {67},
numpages = {13},
location = {Lisbon, Portugal},
series = {ICSE '24}
}

@misc{llm-api-misuse,
      title={Generating API Parameter Security Rules with LLM for API Misuse Detection}, 
      author={Jinghua Liu and Yi Yang and Kai Chen and Miaoqian Lin},
      year={2024},
      eprint={2409.09288v2},
      archivePrefix={arXiv},
      primaryClass={cs.CR},
      url={https://arxiv.org/pdf/2409.09288v2},
}

@inproceedings{llm-code-sum,
author = {Wu, Yifan and Li, Ying and Yu, Siyu},
title = {Commit Message Generation via ChatGPT: How Far Are We?},
year = {2024},
isbn = {9798400706097},
publisher = {Association for Computing Machinery},
address = {New York, NY, USA},
url = {https://doi.org/10.1145/3650105.3652300},
doi = {10.1145/3650105.3652300},
booktitle = {Proceedings of the 2024 IEEE/ACM First International Conference on AI Foundation Models and Software Engineering},
pages = {124–129},
numpages = {6},
location = {Lisbon, Portugal},
series = {FORGE '24}
}

@inproceedings{llm-apr,
author = {Wu, Yonghao and Li, Zheng and Zhang, Jie M. and Liu, Yong},
title = {ConDefects: A Complementary Dataset to Address the Data Leakage Concern for LLM-Based Fault Localization and Program Repair},
year = {2024},
isbn = {9798400706585},
publisher = {Association for Computing Machinery},
address = {New York, NY, USA},
url = {https://doi.org/10.1145/3663529.3663815},
doi = {10.1145/3663529.3663815},
booktitle = {Companion Proceedings of the 32nd ACM International Conference on the Foundations of Software Engineering},
pages = {642–646},
numpages = {5},
location = {Porto de Galinhas, Brazil},
series = {FSE 2024}
}

@misc{safe,
      title={Automated Proof Generation for Rust Code via Self-Evolution}, 
      author={Tianyu Chen and Shuai Lu and Shan Lu and Yeyun Gong and Chenyuan Yang and Xuheng Li and Md Rakib Hossain Misu and Hao Yu and Nan Duan and Peng Cheng and Fan Yang and Shuvendu K Lahiri and Tao Xie and Lidong Zhou},
      year={2024},
      eprint={2410.15756},
      archivePrefix={arXiv},
      primaryClass={cs.SE},
      url={https://arxiv.org/abs/2410.15756}, 
}

@misc{jiannan-yao-proof-syntheis,
      title={Leveraging Large Language Models for Automated Proof Synthesis in Rust}, 
      author={Jianan Yao and Ziqiao Zhou and Weiteng Chen and Weidong Cui},
      year={2023},
      eprint={2311.03739},
      archivePrefix={arXiv},
      primaryClass={cs.FL},
      url={https://arxiv.org/abs/2311.03739}, 
}

@misc{autoverus,
      title={AutoVerus: Automated Proof Generation for Rust Code}, 
      author={Chenyuan Yang and Xuheng Li and Md Rakib Hossain Misu and Jianan Yao and Weidong Cui and Yeyun Gong and Chris Hawblitzel and Shuvendu Lahiri and Jacob R. Lorch and Shuai Lu and Fan Yang and Ziqiao Zhou and Shan Lu},
      year={2025},
      eprint={2409.13082},
      archivePrefix={arXiv},
      primaryClass={cs.SE},
      url={https://arxiv.org/abs/2409.13082}, 
}

@inproceedings{rulf,
author = {Jiang, Jianfeng and Xu, Hui and Zhou, Yangfan},
title = {RULF: rust library fuzzing via API dependency graph traversal},
year = {2022},
isbn = {9781665403375},
publisher = {IEEE Press},
url = {https://doi.org/10.1109/ASE51524.2021.9678813},
doi = {10.1109/ASE51524.2021.9678813},
booktitle = {Proceedings of the 36th IEEE/ACM International Conference on Automated Software Engineering},
pages = {581–592},
numpages = {12},
address = {Melbourne, Australia},
series = {ASE '21},
}

@inproceedings{rpg,
author = {Xu, Zhiwu and Wu, Bohao and Wen, Cheng and Zhang, Bin and Qin, Shengchao and He, Mengda},
title = {RPG: Rust Library Fuzzing with Pool-based Fuzz Target Generation and Generic Support},
year = {2024},
isbn = {9798400702174},
publisher = {Association for Computing Machinery},
address = {New York, NY, USA},
url = {https://doi.org/10.1145/3597503.3639102},
doi = {10.1145/3597503.3639102},
booktitle = {Proceedings of the IEEE/ACM 46th International Conference on Software Engineering},
articleno = {124},
numpages = {13},
location = {Lisbon, Portugal},
series = {ICSE '24}
}

@article{mem-safety-challenge,
author = {Xu, Hui and Chen, Zhuangbin and Sun, Mingshen and Zhou, Yangfan and Lyu, Michael R.},
title = {Memory-Safety Challenge Considered Solved? An In-Depth Study with All Rust CVEs},
year = {2021},
issue_date = {January 2022},
publisher = {Association for Computing Machinery},
address = {New York, NY, USA},
volume = {31},
number = {1},
issn = {1049-331X},
url = {https://doi.org/10.1145/3466642},
doi = {10.1145/3466642},
journal = {ACM Trans. Softw. Eng. Methodol.},
month = sep,
articleno = {3},
numpages = {25}
}

@inproceedings{mem-thread-safety,
author = {Qin, Boqin and Chen, Yilun and Yu, Zeming and Song, Linhai and Zhang, Yiying},
title = {Understanding memory and thread safety practices and issues in real-world Rust programs},
year = {2020},
isbn = {9781450376136},
publisher = {Association for Computing Machinery},
address = {New York, NY, USA},
url = {https://doi.org/10.1145/3385412.3386036},
doi = {10.1145/3385412.3386036},
booktitle = {Proceedings of the 41st ACM SIGPLAN Conference on Programming Language Design and Implementation},
pages = {763–779},
numpages = {17},
location = {London, UK},
series = {PLDI 2020}
}

@inproceedings{unsafe-interview,
author = {H\"{o}ltervennhoff, Sandra and Klostermeyer, Philip and W\"{o}hler, Noah and Acar, Yasemin and Fahl, Sascha},
title = {"I wouldn't want my unsafe code to run my pacemaker": an interview study on the use, comprehension, and perceived risks of unsafe rust},
year = {2023},
isbn = {978-1-939133-37-3},
publisher = {USENIX Association},
address = {USA},
booktitle = {Proceedings of the 32nd USENIX Conference on Security Symposium},
articleno = {141},
numpages = {17},
location = {Anaheim, CA, USA},
series = {SEC '23}
}

@misc{crates-io,
    author = "The crates.io Team",
    title = "The Rust community's crate registry",
    year =  "2025",
    note = "Last accessed Nov. 2025", 
    url = "https://crates.io"
}

@misc{kani-firecracker,
    title = "Using Kani to Validate Security Boundaries in AWS Firecracker",
    author = "Kani Verifier",
    url = "https://model-checking.github.io/kani-verifier-blog/2023/08/31/using-kani-to-validate-security-boundaries-in-aws-firecracker.html",
    year =  "2023",
    month = aug,
}

@misc{kani-hifitime,
    title = "How Kani helped find bugs in Hifitime",
    author = "Kani Verifier",
    url = "https://model-checking.github.io/kani-verifier-blog/2023/03/31/how-kani-helped-find-bugs-in-hifitime.html",
  year =  "2023",
    month = may,
}

@misc{kani-s2n-quic,
    title = "Use Kani action in CI",
    author = "Kani Verifier",
    url = "https://github.com/aws/s2n-quic/pull/1556",
    year =  "2022",
    month = oct,
}

@misc{zero-shot,
      title={Finetuned Language Models Are Zero-Shot Learners}, 
      author={Jason Wei and Maarten Bosma and Vincent Y. Zhao and Kelvin Guu and Adams Wei Yu and Brian Lester and Nan Du and Andrew M. Dai and Quoc V. Le},
      year={2022},
      eprint={2109.01652},
      archivePrefix={arXiv},
      primaryClass={cs.CL},
      url={https://arxiv.org/abs/2109.01652}, 
}

@inproceedings{few-shot,
author = {Brown, Tom B. and Mann, Benjamin and Ryder, Nick and Subbiah, Melanie and Kaplan, Jared and Dhariwal, Prafulla and Neelakantan, Arvind and Shyam, Pranav and Sastry, Girish and Askell, Amanda and Agarwal, Sandhini and Herbert-Voss, Ariel and Krueger, Gretchen and Henighan, Tom and Child, Rewon and Ramesh, Aditya and Ziegler, Daniel M. and Wu, Jeffrey and Winter, Clemens and Hesse, Christopher and Chen, Mark and Sigler, Eric and Litwin, Mateusz and Gray, Scott and Chess, Benjamin and Clark, Jack and Berner, Christopher and McCandlish, Sam and Radford, Alec and Sutskever, Ilya and Amodei, Dario},
title = {Language models are few-shot learners},
year = {2020},
isbn = {9781713829546},
publisher = {Curran Associates Inc.},
address = {Red Hook, NY, USA},
booktitle = {Proceedings of the 34th International Conference on Neural Information Processing Systems},
articleno = {159},
numpages = {25},
location = {Vancouver, BC, Canada},
series = {NIPS '20}
}

@inproceedings{cot,
author = {Wei, Jason and Wang, Xuezhi and Schuurmans, Dale and Bosma, Maarten and Ichter, Brian and Xia, Fei and Chi, Ed H. and Le, Quoc V. and Zhou, Denny},
title = {Chain-of-thought prompting elicits reasoning in large language models},
year = {2022},
isbn = {9781713871088},
publisher = {Curran Associates Inc.},
address = {Red Hook, NY, USA},
booktitle = {Proceedings of the 36th International Conference on Neural Information Processing Systems},
articleno = {1800},
numpages = {14},
location = {New Orleans, LA, USA},
series = {NIPS '22}
}

@misc{react,
      title={ReAct: Synergizing Reasoning and Acting in Language Models}, 
      author={Shunyu Yao and Jeffrey Zhao and Dian Yu and Nan Du and Izhak Shafran and Karthik Narasimhan and Yuan Cao},
      year={2023},
      eprint={2210.03629},
      archivePrefix={arXiv},
      primaryClass={cs.CL},
      url={https://arxiv.org/abs/2210.03629}, 
}

@misc{prompt-chaining,
author = {Saravia, Elvis},
journal = {https://github.com/dair-ai/Prompt-Engineering-Guide},
month = {12},
title = {Prompt Engineering Guide},
year = {2022}
}

@misc{gpt-4.1-2025-04-14,
    title = {GPT-4.1},
    author = "OpenAI",
    url = {https://platform.openai.com/docs/models/gpt-4.1},
    year = "2025",
    month = apr
}

@misc{ds-v3,
    title = {deepseek-chat},
    author = "The DeepSeek Team",
    url = {https://api-docs.deepseek.com/},
    year = "2025",
    month = may
}

@misc{ds-r1,
    title = {deepseek-reasoner},
    author = "The DeepSeek Team",
    url = {https://api-docs.deepseek.com/},
    year = "2025",
    month = may
}

@misc{claude-4,
    title = {Claude Sonnet 4},
    author = "Anthropic",
    url = {https://docs.anthropic.com/en/docs/about-claude/models/overview#model-comparison-table},
    year = "2025",
    month = may
}

@misc{rvt,
    title = {"Rust Verification Tools"},
    author = {"The RVT developers"},
    url = {https://project-oak.github.io/rust-verification-tools/about.html},
    year = "2025", 
    note = "Last accessed Nov. 2025"
}

@misc{autoharness,
    title = "Autoharness",
    author = "Kani Developers", 
    year = "2025",
    month = Jul,
    url = {https://model-checking.github.io/kani/reference/experimental/autoharness.html}
}

@InProceedings{ermedahl2007loop,
  author =	{Ermedahl, Andreas and Sandberg, Christer and Gustafsson, Jan and Bygde, Stefan and Lisper, Bj\"{o}rn},
  title =	{{Loop Bound Analysis based on a Combination of Program Slicing, Abstract Interpretation, and Invariant Analysis}},
  booktitle =	{7th International Workshop on Worst-Case Execution Time Analysis (WCET'07)},
  pages =	{1--6},
  series =	{Open Access Series in Informatics (OASIcs)},
  ISBN =	{978-3-939897-05-7},
  ISSN =	{2190-6807},
  year =	{2007},
  volume =	{6},
  editor =	{Rochange, Christine},
  publisher =	{Schloss Dagstuhl -- Leibniz-Zentrum f{\"u}r Informatik},
  address =	{Dagstuhl, Germany},
  URL =		{https://drops.dagstuhl.de/entities/document/10.4230/OASIcs.WCET.2007.1194},
  URN =		{urn:nbn:de:0030-drops-11946},
  doi =		{10.4230/OASIcs.WCET.2007.1194}
}
